%% file: bare_jrnl_compsoc.tex
\documentclass[10pt,journal,cspaper,compsoc]{IEEEtran}
\ifCLASSOPTIONcompsoc
  \usepackage[nocompress]{cite}
\else
  \usepackage{cite}
\fi
\ifCLASSINFOpdf
  \usepackage[pdftex]{graphicx}
\else
  \usepackage[dvips]{graphicx}
\fi
\usepackage[cmex10]{amsmath}
\usepackage{amssymb}
\usepackage{algorithmic}

\usepackage{array}

\usepackage{mdwmath}
\usepackage{mdwtab}

\usepackage{eqparbox}
\usepackage{graphicx,wrapfig,epsfig}
\usepackage{subcaption}

\usepackage{fixltx2e}

\usepackage{stfloats}

\ifCLASSOPTIONcaptionsoff
  \usepackage[nomarkers]{endfloat}
 \let\MYoriglatexcaption\caption
 \renewcommand{\caption}[2][\relax]{\MYoriglatexcaption[#2]{#2}}
\fi
\usepackage{url}

\usepackage{slashbox}
\usepackage{algorithm2e}
\usepackage{framed}
\usepackage{multirow}
\usepackage{tabularx}
\usepackage[export]{adjustbox}

\begin{document}
\newcommand{\subjectto}{\mathop{\textrm{subject to}}}
\newcommand{\argmax}{\mathop{\textrm{argmax}}}
\newcommand{\argmin}{\mathop{\textrm{argmin}}}
\newcommand{\shortestpath}{\mathop{\textrm{shortest path}}}
\newtheorem{myDefinition}{Definition}
\newtheorem{myTheorem}{Theorem}

%
\title{Nonlinear Dimensionality Reduction via Path-Based Isometric Mapping}
%
%
%
%

\author{Amir~Najafi,
        Amir~Joudaki,
        and Emad~Fatemizadeh
\IEEEcompsocitemizethanks{\IEEEcompsocthanksitem Authors are affiliated with the Biomedical Signal and Image Processing Laboratory (BiSIPL), Department
of Electrical Engineering, Sharif University of Technology, Tehran,
Iran.\protect\\
E-mails: najafi@ee.sharif.edu, amir.judaki@gmail.com, fatemizadeh@sharif.edu}
\thanks{}}

%
%

\markboth{IEEE Transactions on Pattern Analysis and Machine Intelligence}
{Shell \MakeLowercase{\textit{et al.}}: Bare Demo of IEEEtran.cls for Computer Society Journals}
%


\IEEEcompsoctitleabstractindextext{%
\begin{abstract}
Nonlinear dimensionality reduction methods have demonstrated top-notch performance in many pattern recognition and image classification tasks. Despite their popularity, they suffer from highly expensive time and memory requirements, which render them inapplicable to large-scale datasets. To leverage such cases we propose a new method called ``Path-Based Isomap". Similar to Isomap, we exploit geodesic paths to find the low-dimensional embedding. However, instead of preserving pairwise geodesic distances, the low-dimensional embedding is computed via a path-mapping algorithm. Due to the much fewer number of paths compared to number of data points, a significant improvement in time and memory complexity without any decline in performance is achieved. The method demonstrates state-of-the-art performance on well-known synthetic and real-world datasets, as well as in the presence of noise.
\end{abstract}

}

\maketitle

\IEEEdisplaynotcompsoctitleabstractindextext

%
\IEEEpeerreviewmaketitle

\section{Introduction}
%
%

%
%
%
%
\IEEEPARstart{O}{ne} of the fundamental problems in machine learning and pattern recognition is to discover compact representations of high-dimensional data. The need to analyze and visualize multivariate data has yielded a surge of interest in dimensionality reduction research \cite{demartines1997curvilinear}, \cite{hinton2002stochastic}, \cite{donoho2003hessian}, \cite{lespinats2009rankvisu}. In particular, manifold learning techniques such as Isomap \cite{tenenbaum2000global}, Locally Linear Embedding (LLE) \cite{roweis2000nonlinear}, and Laplacian Eigenmaps \cite{belkin2003laplacian} have outperformed classical methods like Principal Component Analysis (PCA) and Multi-Dimensional Scaling (MDS) \cite{borg2005modern} in harnessing non-linear data structures \cite{timofte2012iterative}, \cite{guo2011simultaneous}. However, their high time and memory complexity impose severe limitations on their scalabality \cite{kohonen2001self}. To overcome this drawback, we set out to develop a method with lower computational costs yet the same performance. 

Throughout the paper it is assumed that data samples lie on a smooth low-dimensional manifold \cite{baraniuk2009random}, \cite{hegde2012learning}. In the first stage, these data samples are covered by a set of geodesic paths, resulting in a network of intersecting routes. The main point is that data samples that belong to a geodesic path approximately lie on a straight line in the compact representation \cite{bernstein2000graph}. Thus a mapping scheme is developed to compute the lines in the destination space. The scheme is formulated as an optimization problem that attempts to preserve topology of the network of paths instead of pairwise geodesic distances. This is a crucial difference between our approach and Isomap that yields remarkable cost savings.‬

Experiments on commonly used synthetic and real-world datasets substantiates superiority of our method in terms of efficiency. They also demonstrate that this achievement do not come at the cost of performance, stability, or robustness of the algorithm. Another advantage of this algorithm is that the aforementioned optimization problem has an analytical solution, that avoids local minima and has deterministic time bounds.

Rest of the paper is organized as follows: Section 2 overviews the Isomap algorithm. Section 3 highlights the main idea of the paper. Section 4 gives a stochastic algorithm for covering the data samples with a set of geodesic paths. Section 5 details the path-mapping scheme. Section 6 is dedicated to complexity analysis of the algorithm. Section 7 discusses experimental results, and finally the conclusions are made in section 8.

%

\section{Isometric Mapping (Isomap)}
This section briefly explains the Isomap algorithm. Assume a cloud of high dimensional data points $\left\{{\bf x}_1,{\bf x}_2,...,{\bf x}_N\right\},{\bf x}_i \in \mathcal{R}^M$ lie on a smooth $K$-dimensional manifold. In most cases of practical interest $K$ is  much smaller than the data dimension $M$ ($K\ll M$). Isomap builds upon MDS but attempts to compute the low-dimensional representation by estimating pairwise geodesic distances.

For sufficiently close pairs, referred to as neighboring points, the euclidean distance provides a good approximation of geodesic distance \cite{bernstein2000graph}, \cite{balasubramanian2002isomap}. For faraway points, one needs to walk through these neighboring pairs in the shortest way possible to evaluate the geodesic distance. That can be achieved efficiently by applying a shortest path algorithm on a graph comprising edges that connect neighboring points.

Here we introduce notations for these concepts. The graph is represented as $G=(V,E)$ in which $V=\left\{{\bf x}_1,{\bf x}_2,...,{\bf x}_N\right\}$ denotes the set of nodes, and $E$ is the set of edges connecting neighboring samples. Two ways determining the neighbors of a point are K-nearest neighbors \cite{seidl1998optimal}, or all points within a fixed range $\epsilon$. In this paper we utilize the former method. For neighboring nodes ${\bf x}_i$ and ${\bf x}_j$ the weight is taken to be $w_{i,j} = ||{\bf x}_i-{\bf x}_j||_2$. If we take ${\bf x}_i \leadsto {\bf x}_j$ to be the shortest route between ${\bf x}_i$ and ${\bf x}_j$, we could compute geodesic distances as $d^G({\bf x}_i,{\bf x}_j) = w({\bf x}_i \leadsto {\bf x}_j)$ in which $w(.)$ denotes weight of the path.

Finally, we seek a set of low-dimensional points denoted by $\left\{{\bf y}_1,{\bf y}_2,...,{\bf y}_N\right\}$ in $\mathcal{R}^K$ that preserves pairwise geodesic distances. This can be accomplished via a classical MDS approach.


\section{A Path-Based Approach}
Isomap discards the fact that a shortest path $\mathcal{L}$ will be approximately mapped to a straight line in the representational space. In this regard, if we enforce each ${\bf y}_i \in \mathcal{L}$ to lie exactly on a straight line, degrees of freedom will be reduced dramatically. Here is the explanation behind this fact: assume ${\bf x}_j$ to lie on the shortest path between ${\bf x}_i$ and ${\bf x}_k$ denoted as $\mathcal{L}={\bf x}_i \leadsto {\bf x}_j \leadsto {\bf x}_k$. It can be concluded that $d^G({\bf x}_i,{\bf x}_j)+d^G({\bf x}_j,{\bf x}_k)$ equals to $d^G({\bf x}_i,{\bf x}_k)$. Since MDS tries to preserves these geodesic distances in the representational space, it attempts to satisfy the equation $||{\bf y}_i-{\bf y}_j||+ ||{\bf y}_j-{\bf y}_k||=||{\bf y}_i-{\bf y}_k||$. This, in turn, implies that the three points must lie on a straight line. Not to mention this is the ideal case, without any noise and in the limit of infinite samples. Regarding the fact that points were chosen arbitrarily, it can be concluded that all points on a shortest path must lie on a straight line. 

\begin{figure}[t]
\centering
        \begin{subfigure}[b]{0.22\textwidth}
                \includegraphics[trim=0.7in 0.4in 0.6in 0.3in,clip,width=\textwidth]{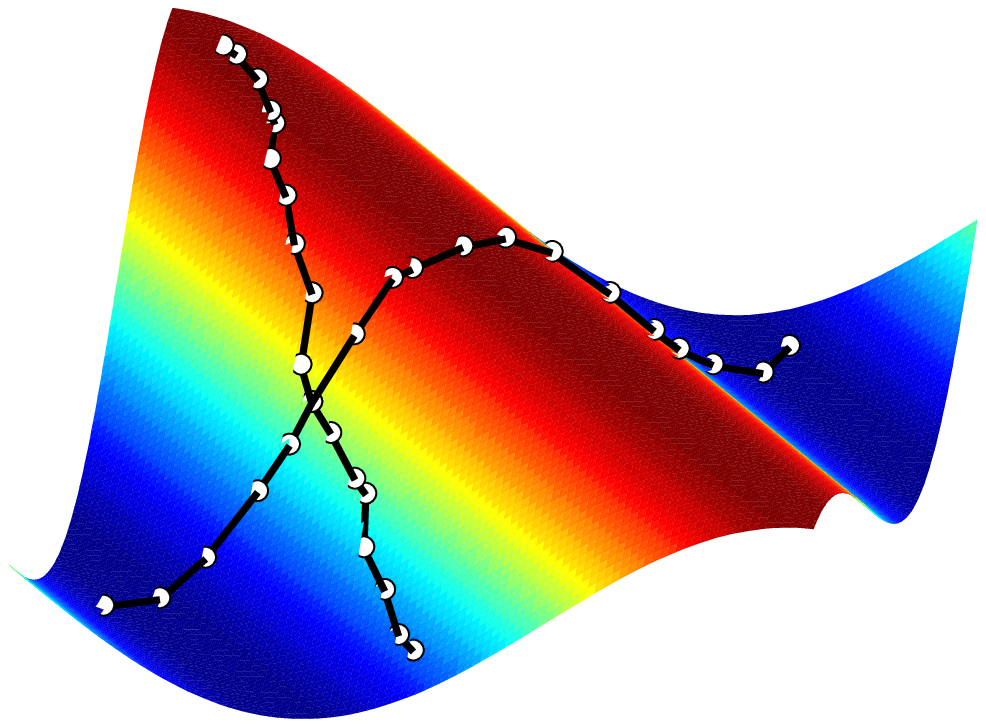}
        \end{subfigure}%
        ~
        \begin{subfigure}[b]{0.22\textwidth}
                \includegraphics[trim=0.6in 0.2in 0.4in 0.3in,clip,width=\textwidth]{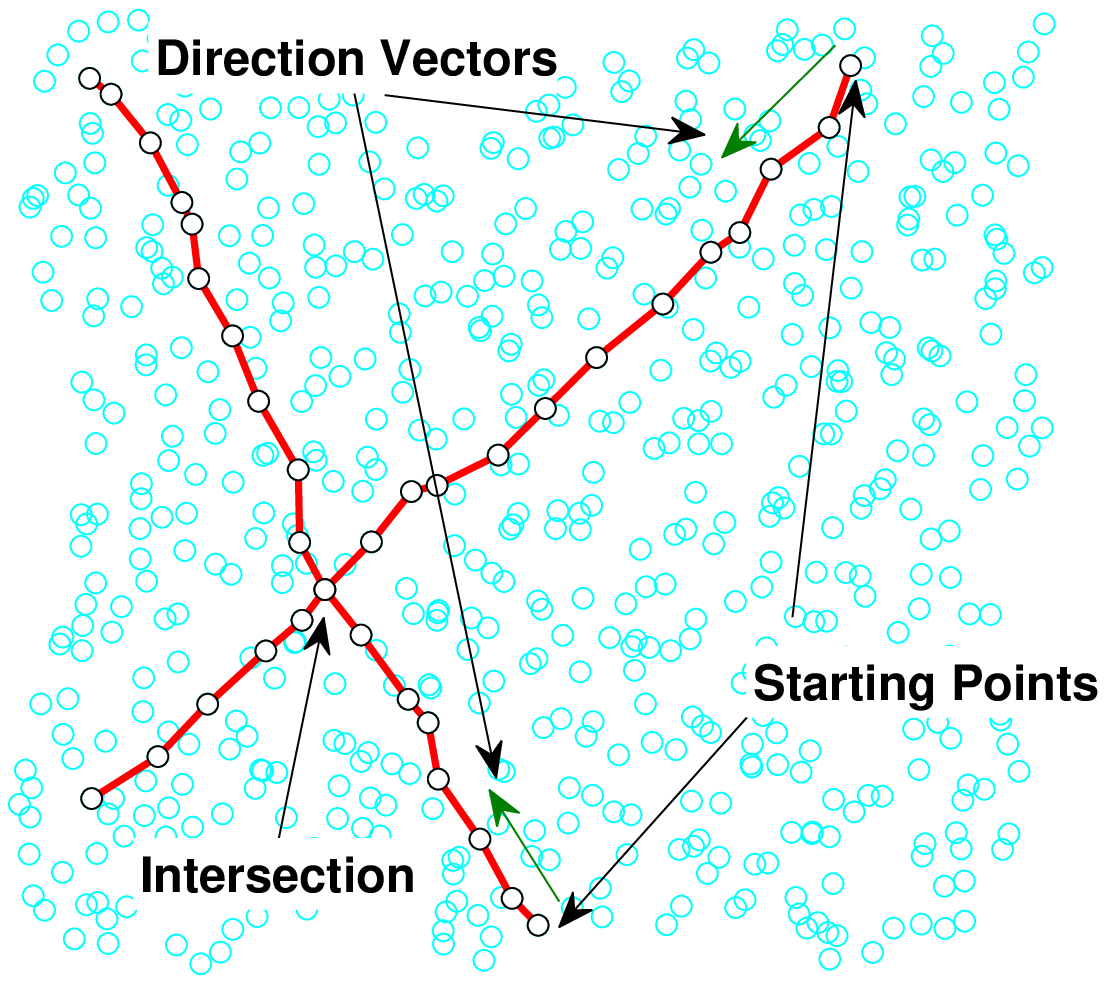}
        \end{subfigure}%
        \caption{{\small Graphical illustration of the scheme that maps geodesic paths on the manifold to straight lines. Direction and starting point of each line are tuned in a way that global geometry of data points is preserved.}}\label{fig_1_1}
\end{figure}

Thus, assuming $L$ to be the number of points in ${\bf x}_i \leadsto {\bf x}_k$, the number of degrees of freedom drops from $L K$, describing $L$ points in $\mathcal{R}^K$, to $2 K$, describing the starting point and direction of a line in $\mathcal{R}^K$. Suppose starting point ${\bf x}_{\alpha} \in \mathcal{L}$ is mapped to ${\bf y}_{\alpha} \in \mathcal{R}^K$ and the direction of the straight line, $\hat{\boldsymbol{v}}$, is discovered. Any other point on the path ${\bf x}_{\beta} \in \mathcal{L}$ can be mapped automatically: 
\begin{equation}
\label{eq1}
{\bf y}_{\beta} = {\bf y}_{\alpha}+d^G({\bf x}_{\alpha},{\bf x}_{\beta})\hat{\boldsymbol{v}}
\end{equation}
which incorporates the fact that $d^G({\bf x}_{\alpha},{\bf x}_{\beta})=||{\bf y}_{\alpha}-{\bf y}_{\beta}||$. This approach is sketched out in Fig.\ref{fig_1_1}. As can be seen two geodesic paths on a manifold in $\mathcal{R}^3$ are approximately mapped to straight lines in $\mathcal{R}^2$.

There still remains two issues to be addressed. First is to find a set of shortest paths $\Omega =\left\{ \mathcal{L}_1, ... , \mathcal{L}_P\right\}$ that will cover all nodes of the graph, and second is to develop a scheme that maps shortest paths in $\mathcal{R}^M$ to straight lines in $\mathcal{R}^K$. These issues are discussed respectively in sections 4 and 5. 

\section{Stochastic Shortest Path Covering}

\begin{algorithm}[b] 
 \TitleOfAlgo{Stochastic Shortest Path Covering (SSPC)}
 \begin{framed}
 \SetAlgoLined 
 \KwData{$R\leftarrow V = \left\{{\bf x}_1,{\bf x}_2,...,{\bf x}_N\right\}$\\
 		 \hspace{9.5mm}$\Omega\leftarrow\emptyset$}  
 \While{$R\neq\emptyset$}
 {\vspace{1mm}
 Choose a random node in $R$, denoted as $s\in R$.\vspace{1mm}\\
 Compute all the shortest paths that start from $s$ and end to other members of $R$.\vspace{1mm}\\
 Find the path that overlaps the most with $R$, denoted as $\mathcal{L}^*$.\vspace{1mm}\\
 $\Omega\leftarrow\Omega\cup\mathcal{L}^*$\vspace{1mm}\\
 $R\leftarrow R-\mathcal{L}^*$\vspace{1mm}
  }  
  \KwResult{$\Omega$ = A sufficient set of shortest paths.  }
 \end{framed}
 \vspace{-3mm}
\end{algorithm}

This section presents an stochastic algorithm for covering the graph with a set of shortest paths called "Stochastic Shortest Path Covering (SSPC)".

The general problem of graph covering via shortest paths is known to be NP-hard \cite{boothe2007graph}. Thus we set out to find a sub-optimal solution with a stochastic approach. In practice our sub-optimal algorithm yields substantial time and space savings.
 
\begin{figure*}[t]
        \begin{subfigure}[b]{0.29\textwidth}
                \includegraphics[trim=1.75in 1.3in 1.75in 1.35in,clip,width=\textwidth]{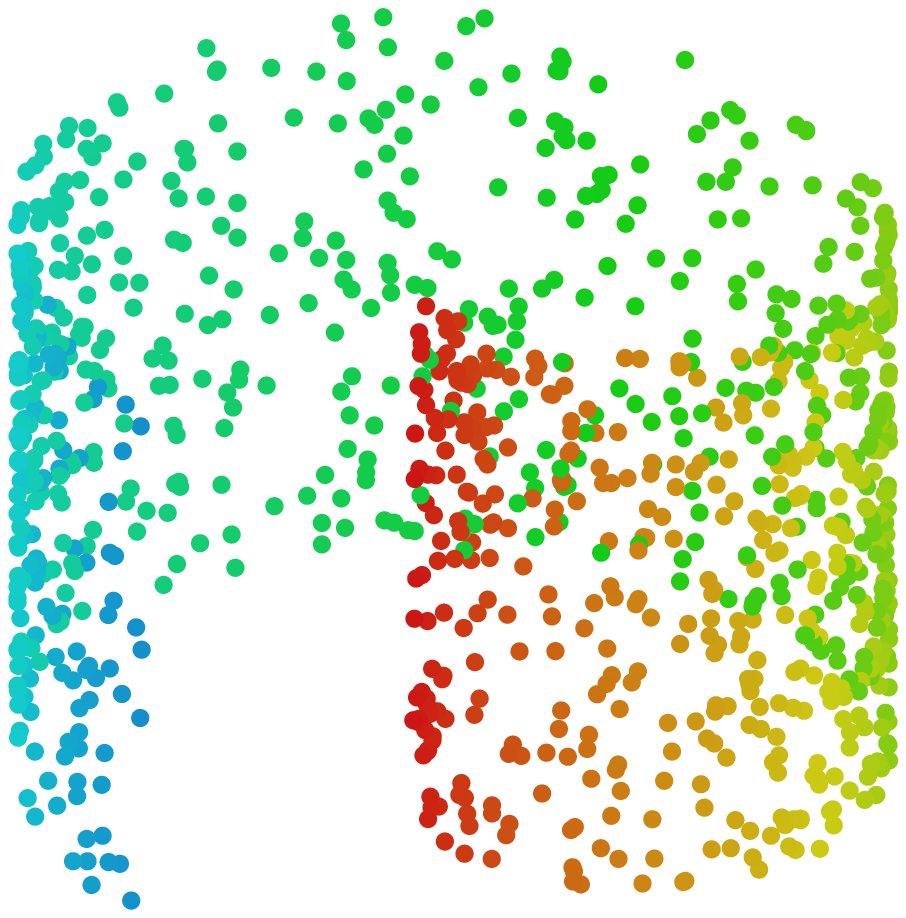}
                \caption{}
        \end{subfigure}%
        ~
        \begin{subfigure}[b]{0.05\textwidth}
		        \includegraphics[trim=-4.1in 0.05in 0.3in 0.5in,clip,angle=90,width=\textwidth]{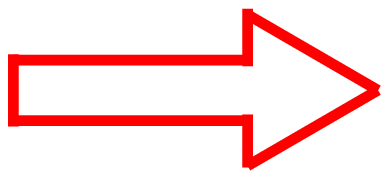}	
        \end{subfigure}%
        ~
        \begin{subfigure}[b]{0.29\textwidth}
                \includegraphics[trim=1.75in 1.3in 1.75in 1.35in,clip,width=\textwidth]{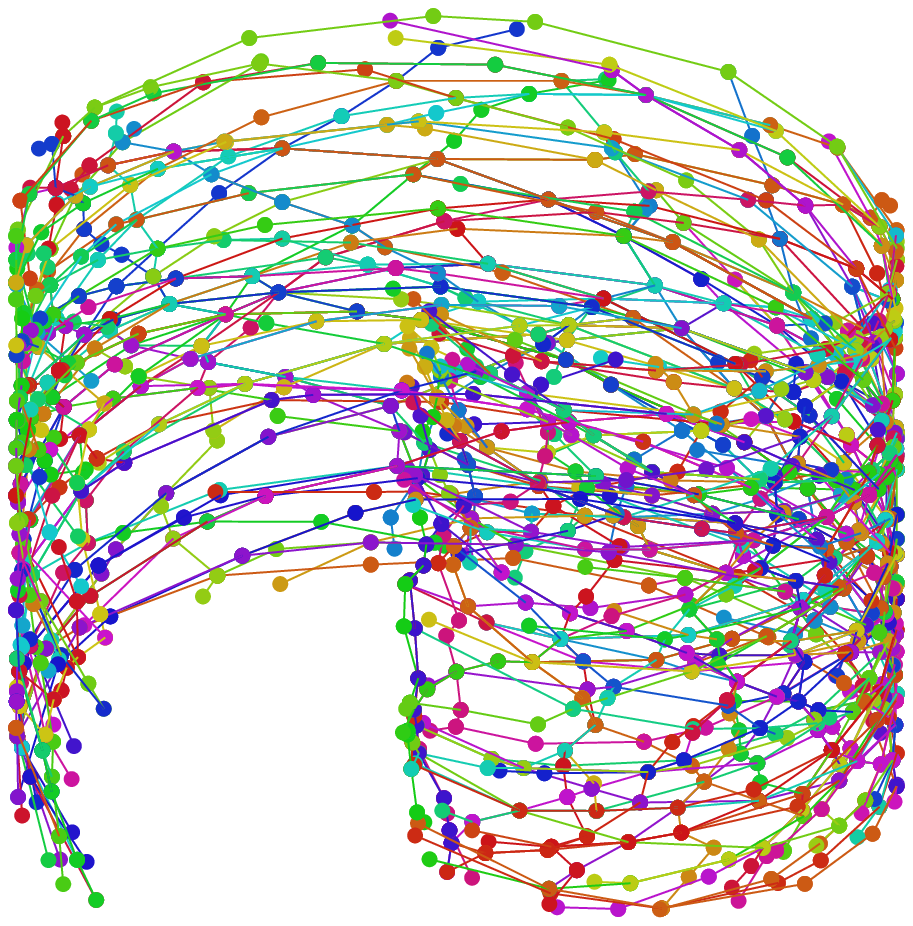}
                \caption{}
        \end{subfigure}%
        ~
        \begin{subfigure}[b]{0.05\textwidth}
		        \includegraphics[trim=-4.1in 0.05in 0.3in 0.5in,clip,angle=90,width=\textwidth]{arrow.eps}
        \end{subfigure}%
        ~
        \begin{subfigure}[b]{0.29\textwidth}
                \includegraphics[trim=0.6in 0.2in 0.5in 0.3in,clip,width=\textwidth]{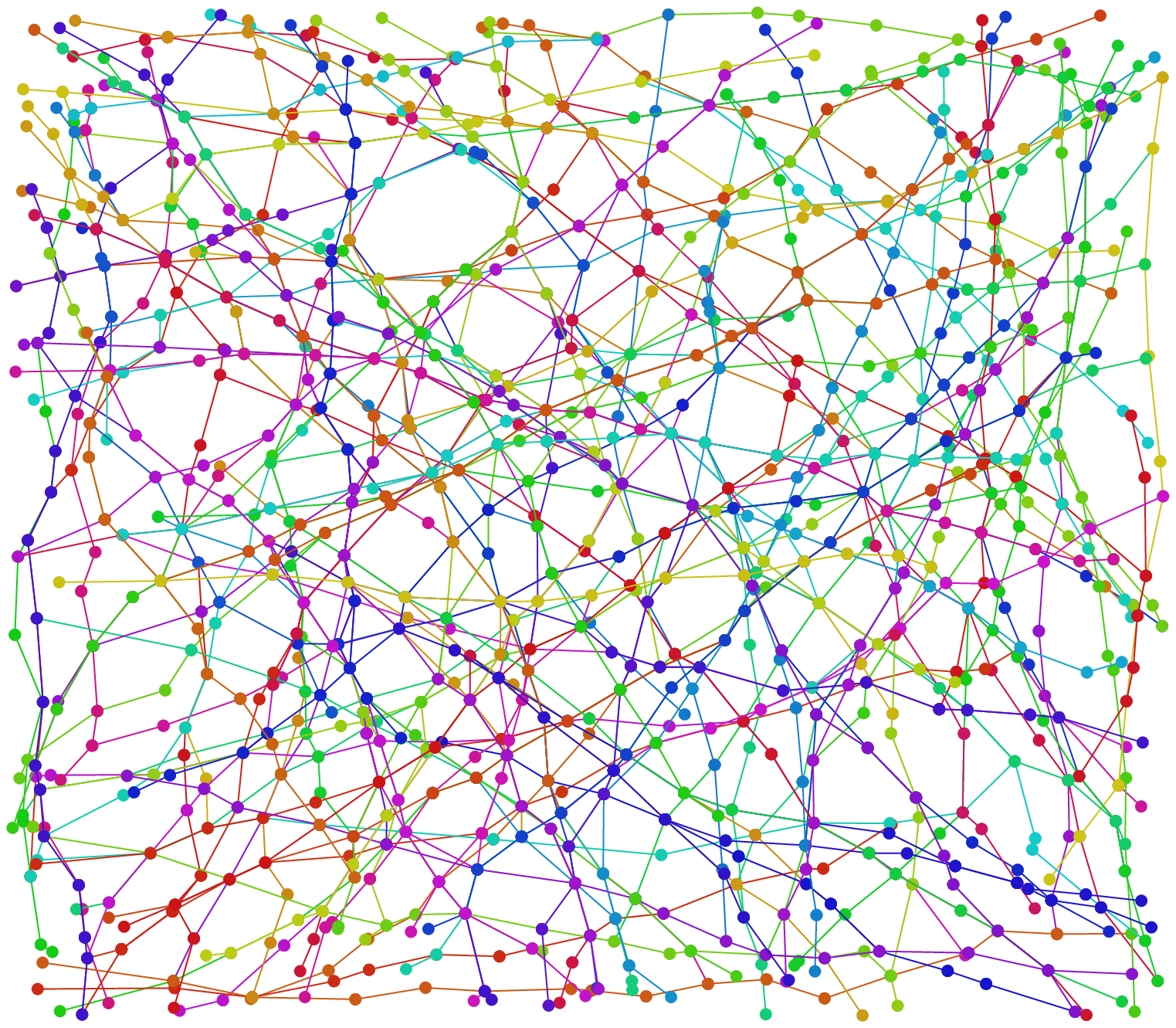}
                \caption{}
        \end{subfigure}
        \vspace{1mm}
        \caption{{\small SSPC sample run on a Swiss-Roll dataset consisting of $1000$ data points. The number of paths obtained by the method in this example is $P=154$. (a) The high-dimensional data in $\mathcal{R}^3$. (b) Result of the SSPC in $\mathcal{R}^3$. (c) The results shown for the unfolded manifold.}}\label{fig_4_1}
\end{figure*}

The general idea is to iteratively cover as many nodes as possible. First we initialize the set of uncovered points, denoted by $R$, to include all vertices of $V$. In each step, a source node $s \in R$ is selected randomly. Among shortest paths starting from $s$, the path $\mathcal{L}^*$, which overlaps the most with $R$ is chosen. The nodes of $\mathcal{L}^*$ are deducted from $R$. We repeat this procedure until there is no point uncovered, i.e. $R=\emptyset$. This results in a stable network of intersecting paths. A pseudo-code of the method is presented. Fig.\ref{fig_4_1} shows an example of applying the SSPC on Swiss-Roll dataset.

\begin{figure}[t]
\centering
    \includegraphics[trim=0.5in 0in 0in 0.3in,clip,width=0.45\textwidth]{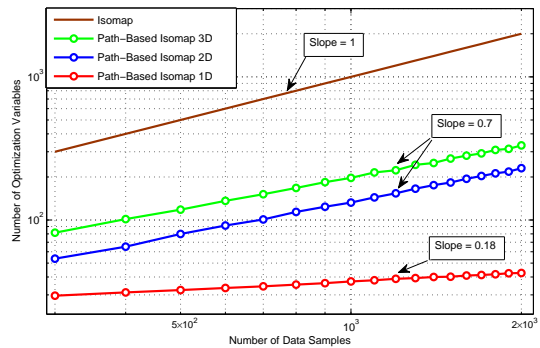}
    \caption{{\small Number of optimization variables depicted as a function of number of data points in a log-log plot. Reduction in complexity is shown for $3$ manifold dimensionalities. The original number of variables used in Isomap is also plotted for comparison.}}
    \label{fig_4_2}
\end{figure}

As it is evident in Fig.\ref{fig_4_1} the number of paths collected by SSPC is significantly fewer than the number of samples. We have empirically investigated the extent of complexity reduction by SSPC algorithm on a number of datasets. It turns out that its contribution to complexity reduction largely depends on the inherent manifold dimensionality. Fig.\ref{fig_4_2} shows number of optimization variables (degrees of freedom) versus number of data points for three manifold dimensionalities: $K=1, 2,$ and $3$. For each $K$ the number of optimization variables are averaged over various synthetic datasets. For the sake of comparison the non-reduced number of variables is shown with a slope of 1. The linear dependence in logarithmic scale implies that there is a power-law relation between the two variables:
\begin{equation}
\label{eq8}
P=\alpha N^\gamma
\end{equation}
where $N$ and $P$ are the number of data points and the number of paths respectively. 

The values that fit the data are given in TABLE 1. Experiments demonstrated that for a specific $K$ variations in the resulting $\alpha$ and $\gamma$ were negligible over a variety of datasets. The numbers in the table suggest that for low-dimensional manifolds and large scale datasets complexity reduction is remarkable, whereas for high-dimensional small-scale ones, the improvement gradually fades out. Surprisingly the same exponent $\gamma$ is obtained for $K=2$ and $K=3$. However, it is possible that this observation is due to the curse of dimesniaolity and might not hold for much larger values of $N$.

\begin{table}[b]
\caption{Experimental Parameters}
\begin{center}
\begin{tabular}{| c | c | c | c |}
\hline
\backslashbox[20mm]{Parameters}{$K$} & $1$ & $2$ & $3$ \\ 
\hline\hline 
$\alpha$ & $10.8$ & $1.35$ & $2.40$\\ \hline
$\gamma$ & $0.18$ & $0.69$ & $0.70$\\ \hline
\end{tabular}
\end{center}
\label{tableExp}
\end{table}

\begin{figure}[b]
\centering
    \includegraphics[trim=0.4in 0in 0.3in 0.28in,clip,width=0.45\textwidth]{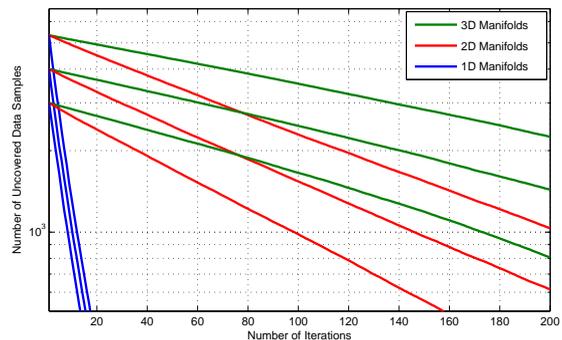}
    \caption{{\small Number of the uncovered data samples shown during execution of SSPC algorithm.}}
    \label{fig_5_2}
\end{figure}

\subsection{Covering Rate Analysis of SSPC}
It turns out that number of data samples covered by each iteration of SSPC gradually decreases during the execution. Moreover, the rate of decay follows an exponential trend. We have seen that the decay exponent is heavily dependent on the innate dimensionality of dataset. Fig.\ref{fig_5_2} shows the number of uncovered samples during the execution of the algorithm for a number of datasets. The linear decline in semi-log plot reveals an exponential decline rate:
\begin{equation}
\label{eq9}
R(n)=Ne^{-\lambda n}
\end{equation}
where $R(n)$ is the number of uncovered samples after the $n$th iteration\footnote{The only exception is that (\ref{eq9}) does not necessarily hold for small values of $R$, such as when $R(n)<100$.}. It is also elucidated by the figure that as long as dimensinality is fixed, the rate of decay is neither sensitive to the embedding geometry nor the initial number of points. In other words, we could express $\lambda$ only in terms of $K$ (manifold dimensionality):
\begin{equation}
\lambda=\lambda{(K)}
\label{eq10}
\end{equation}

In TABLE 2 the exponent of decay ($\lambda$) has been estimated for three values of $K=1, 2$ and $3$. In this regard, SSPC could shed some light on the manifold dimensioality. One could estimate $\lambda$ for an arbitrary dataset and compare it to these values to get a rough estimate of the underlying embedding dimensionality.

The only remaining issue here is the uniqueness of solution for a network of paths computed via SSPC. In other words we should investigate that under which circumstances the path sequences obtained from a graph covering algorithm will uniquely represent a low-dimensional embedding. This issue is addressed in Appendix A, where sufficient conditions are derived and a compensation strategy is proposed.

\section{Parameter Optimization}
What remains to explain is the path-mapping scheme that finds parameters of straight lines in  $\mathcal{R}^K$. One could attempt to estimate initial points and directions of these lines. We approach this problem from an optimization perspective. The cost function will arise from the inherent constraints of the problem. 

Fig.\ref{fig_4_1} clearly shows that the paths resulting from the SSPC algorithm would have numerous intersections. Two paths that cross each other, share a common data sample in a known position according to their starting points and direction vectors. Hence the estimations provided by each path should be close to each other. This lays out the main idea for defining an optimization criteria. This approach is depicted in Fig.\ref{fig_1_1} for a data sample shared between two paths.

\begin{table}[t]
\label{table2}
\caption{Experimental values for $\lambda$}
\begin{center}
\begin{tabular}{| c | c | c | c |}
\hline
$K$ & $1$ & $2$ & $3$\\ \hline
$\lambda$ & $0.123$ & $0.010$ & $0.005$\\ \hline
\end{tabular}
\end{center}
\end{table}

In technical terms, assume that a particular data sample ${\bf y}_\alpha$ is shared by $D$ separate lines $\mathcal{L}_1, \mathcal{L}_2, ..., \mathcal{L}_D$. Each line provides an estimation of the point denoted by $\boldsymbol{P}_1,\boldsymbol{P}_2, ..., \boldsymbol{P}_D$ via line equation (\ref{eq1}). We have:
\begin{gather}
\label{eq10_1}
\left\{
  \begin{array}{l l}
	\boldsymbol{P}_1=\boldsymbol{\xi}_1+l_1\hat{\boldsymbol{v}}_1 \\
	\boldsymbol{P}_2=\boldsymbol{\xi}_2+l_2\hat{\boldsymbol{v}}_2 \\
	\vdots \\
	\boldsymbol{P}_D=\boldsymbol{\xi}_D+l_D\hat{\boldsymbol{v}}_D  
  \end{array} \right.\
\end{gather}
where $\boldsymbol{\xi}_d$ and $\hat{\boldsymbol{v}}_d$ are the starting point and the direction vector of the $d$th line respectively, and $l_d$ is the geodesic distance between the starting sample and ${\bf x}_\alpha$. Ideally, these parameters are tuned in a way that all the estimations become equal. In practice, we attempt to minimize a difference measure among them. A reasonable difference measure can be calculated as:
\begin{gather}
\label{eq10_2}
{\sigma}^2_\alpha=\frac{1}{D}\sum_{d=1}^D||\boldsymbol{P}_d - \bar{\boldsymbol{P}}||^2
\end{gather}
$\bar{\boldsymbol{P}}$ indicates the average of ${\boldsymbol{P}}_d$s. It is clear that a good candidate for cost function will be the sum of difference measures of all shared data points. 

Here we present definitions and terms that will be used later in this section for formulation of cost function:
\begin{equation}
\label{eq11}
\boldsymbol{\xi}_p, \hat{\boldsymbol{v}}_p  \in \mathcal{R}^K,   p=1,2,...,P
\end{equation}
where $\boldsymbol{\xi}_p$ and $\hat{\boldsymbol{v}}_p$ are the starting point and the direction of the $p$th line respectively, and $P$ denotes the total number of paths in $\Omega$. We also designate the $k$th component of the vectors $\boldsymbol{\xi}_p$ and $\hat{\boldsymbol{v}}_p$ by $\xi_p^{(k)}$ and $\hat{v}_p^{(k)}$ respectively. The data samples that are shared among more than one path are numbered by the index $q={1,2,...,Q}$ where $Q$ is the total number of shared samples in $\Omega$. $m_q$ is the number of paths that contain the $q$th shared data sample $(m_q\geq2)$. $\eta_i^{(q)}\in\left\{1,2,...,P\right\}$ is the index of the $i$th line $(i\in\left\{1,2,...,m_q\right\})$ that contains the $q$th shared sample. Finally, $l_i^{(q)}$ is the geodesic distance of this data sample from the starting point of its line.

The optimization problem is formulated as:
\begin{gather}
(\boldsymbol{\xi}^*,\hat{\boldsymbol{v}}^*)=\argmin_{(\boldsymbol{\xi},\hat{\boldsymbol{v}})}\frac{1}{2}\sum_{k=1}^{K}\sum_{q=1}^{Q}\left[\frac{1}{m_q}\sum_{i=1}^{m_q}\left(\xi_{\eta_i^{(q)}}^{(k)}+l_{\eta_i^{(q)}}^{(k)}\hat{v}_{\eta_i^{(q)}}^{(k)}\right)^2\right.
\nonumber \\
\left.-\left(\frac{1}{m_q}\sum_{i=1}^{m_q}\left(\xi_{\eta_i^{(q)}}^{(k)}+l_{\eta_i^{(q)}}^{(k)}\hat{v}_{\eta_i^{(q)}}^{(k)}\right)\right)^2\right] \nonumber 
\end{gather}
\begin{equation}
\label{eq12}
\subjectto \hspace{2mm} \sum_{k=1}^{K}\left(\hat{v}_p^{(k)}\right)^2=1, \hspace{3mm} p=1,2,...,P
\end{equation}
where $\boldsymbol{\xi}$ and $\boldsymbol{\hat{v}}$ are $P\times K$ matrices defined as:
\begin{gather}
\label{eq12_1}
\boldsymbol{\xi} = {\left[\boldsymbol{\xi}_1|\boldsymbol{\xi}_2|\ldots|\boldsymbol{\xi}_P\right]}^T \quad 
\hat{\boldsymbol{v}} = {\left[\hat{\boldsymbol{v}}_1|\hat{\boldsymbol{v}}_2|\ldots|\hat{\boldsymbol{v}}_P\right]}^T
\end{gather}

The constraints in (\ref{eq12}) assure that the direction vectors (rows of $\boldsymbol{\hat{v}}$) have unit norms, keeping the optimization procedure away from finding trivial solutions.

Both the objective function and the constraints in (\ref{eq12}) are quadratic and convex with respect to the line parameters, that means there exists an analytical solution for this problem. By forming the Lagrangian of (\ref{eq12}) and calculating the derivatives with respect to all the variables, it is shown in Appendix D that the optimal direction matrix $\boldsymbol{\hat{v}}^*$ can be obtained from the eigenvectors of the following matrix $\psi$:
\begin{equation}
\label{eq13}
\psi=(B'-A'A^{\dagger}B)
\end{equation}
where $A, B, A'$ and $B'$ are $P\times P$ matrices defined in (\ref{eq14}),(\ref{eq15}),(\ref{eq16}) and (\ref{eq17}) respectively. $A^{\dagger}$ denotes the pseudo-inverse of the matrix $A$ (since $A$ is singular). Eigenvalues of $\psi$ are non-negative since the matrix is positive semi-definite. 

\begin{align}
\label{eq14}
A_{r,s}=\sum_{q=1}^{Q}\sum_{\forall i|\eta_i^{(q)}=r}^{m_q}\left(\frac{1}{m_q}\delta_{r,s}\right)-\sum_{\forall q|\exists i \atop \Rightarrow \eta_i^{(q)}=r}\left(\sum_{\forall i|\eta_i^{(q)}=s}^{m_q}\frac{1}{m_q^2}\right)
\end{align}
\begin{align}
\label{eq15}
B_{r,s}=\sum_{q=1}^{Q}\sum_{\forall i|\eta_i^{(q)}=r}^{m_q}\left(\frac{l_i^{(q)}}{m_q}\delta_{r,s}\right)-\sum_{\forall q|\exists i \atop \Rightarrow \eta_i^{(q)}=r}\left(\sum_{\forall i|\eta_i^{(q)}=s}^{m_q}\frac{l_i^{(q)}}{m_q^2}\right)
\end{align}
And for $A'$ and $B'$ matrices:
\begin{align}
\label{eq16}
A'_{r,s}=\sum_{q=1}^{Q}\sum_{\forall i|\eta_i^{(q)}=r}^{m_q}\left(\frac{l_i^{(q)}}{m_q}\delta_{r,s}\right)-\sum_{\forall q|\exists i=i_0 \atop \Rightarrow \eta_{i_0}^{(q)}=r}\left(\sum_{\forall i|\eta_i^{(q)}=s}^{m_q}\frac{l_{i_0}^{(q)}}{m_q^2}\right)
\end{align}
\begin{align}
\label{eq17}
B'_{r,s}=\sum_{q=1}^{Q}\sum_{\forall i|\atop\eta_i^{(q)}=r}^{m_q}\left(\frac{\left(l_i^{(q)}\right)^2}{m_q}\delta_{r,s}\right)-\nonumber
\\
\sum_{\forall q|\exists i \atop \Rightarrow \eta_i^{(q)}}\left(\sum_{\forall i|\atop\eta_i^{(q)}=s}^{m_q}\frac{\left(l_i^{(q)}l_{i_0}^{(q)}\right)}{m_q^2}\right)
\end{align}
\begin{equation}
r,s\in\left\{1,2,...,P\right\} \nonumber
\end{equation}
where $\delta_{r,s}$ is the Kronecker-delta operator.

Eigenvectors of the matrix $\psi$ which correspond to $K$ smallest positive eigenvalues represent the $P\times K$ direction matrix $\hat{\boldsymbol{v}}^*$. It should be noted that each row must be normalized so the direction vectors would have unit norms. Interestingly, the starting positions matrix $\boldsymbol{\xi}^*$ is linearly related to $\hat{\boldsymbol{v}}^*$:
\begin{equation}
\label{eq18}
\boldsymbol{\xi}^*=-A^{\dagger}B\hat{\boldsymbol{v}}^*
\end{equation}

After solving for the direction vectors $\hat{\boldsymbol{v}}^*$ and corresponding starting points $\boldsymbol{\xi}^*$, we may obtain the low-dimensional representations using an averaging strategy:
\begin{gather}
\label{eq19}
{\bf y}_n=\sum_{p=1}^{L_n}\frac{1}{L_n}\left(\boldsymbol{\xi}^*_{\mu_p^{(n)}}+l_p^{(n)}\hat{\boldsymbol{v}}^*_{\mu_p^{(n)}}\right) \\
n=1,2,...,N \nonumber
\end{gather}
where similar to the notation used before, $L_n$ is total number of the lines that contain the $n$th data sample, $L_n\geq1$ for all $n$. $\mu_p^{(n)}$ is the index of the $p$th line that contains the $n$th data sample and $l_p^{(n)}$ is the length (distance) associated to the mentioned line and data sample.

\section{Computational Complexity Analysis}
In this section we lay out time and memory complexity analysis of the Path-based Isomap and compare it to a number of existing methods. Efficient methods have been proposed for construction of the neighborhood graph $G$ \cite{indyk2004nearest}, \cite{seidl1998optimal}. However, since the procedure is shared among all state-of-the-art methods it is not taken into consideration \cite{van2009dimensionality}. Moreover, it is assumed that number of neighbors, denoted by $\mathfrak{n}$, in K-nearest algorithm is $O(1)$. Since in practice it is not relevant to the number of samples \cite{tenenbaum2000global}, \cite{roweis2000nonlinear}, \cite{belkin2003laplacian}. 

Computation of shortest paths in the SSPC algorithm requires $O(PN\log{N})=O\left(N^{(1+\gamma)}\log{N}\right)$ computations for applying Dijkstra's algorithm \cite{even2011graph} $O(N^{\gamma})$ times. The Singular Value Decomposition (SVD) used in optimization problem requires $O\left(P^3\right)=O\left(N^{(3\gamma)}\right)$ multiplications \cite{henry2010singular}. So the total time complexity of the algorithm is $O\left(N^{\left(1+\gamma\right)}\log{N}+N^{(3\gamma)}\right)$. For memory analysis, there are three major components. First, SVD requires $O\left(N^{\left(2\gamma\right)}\right)$ \cite{henry2010singular}. Second, the $\mathfrak{n}$ neighborhoods for each sample should be saved which, regarding the assumption about $\mathfrak{n}$, needs $O\left(N\right)$. Third and the most important factor is the memory needed to save the shortest paths. In the limit that $N$ goes to infinity, based on isoperimetric inequality \cite{osserman1978isoperimetric}, the average length of these paths would be lower than $O\left(N^{1/K}\right)$. Given the number of paths $O\left(N^{\gamma}\right)$ the memory complexity of this component is $O\left(N^{\left(1/K+\gamma\right)}\right)$. So the total memory complexity is $O\left(N+N^{2\gamma}+N^{\left(1/K+\gamma\right)}\right)$. 

Isomap requires $O\left(N^2\sim N^3\right)$ for computing shortest paths \cite{balasubramanian2002isomap}, \cite{pallottino1984shortest}. Isomap, LLE and Laplacian-Eigenmaps need the SVD of an $N\times N$ matrix in the final stage \cite{tenenbaum2000global}, \cite{roweis2000nonlinear}, \cite{belkin2003laplacian}. Due to the sparsity of this matrix for LLE and Laplacian-Eigenmaps the complexity will be reduced. Hence computations in the latter stage will be $O\left(N^3\right)$ for Isomap, and $O\left(N^2\right)$ for LLE and Laplacian-Eigenmaps \cite{berry1992large}. For memory the only important component, that is the memory required by SVD, is $O\left(N^2\right)$ \cite{berry1992large}.
\section{Experimental Results}
In this section the performance of the proposed algorithm on both synthetic and real-world datasets has been simulated.

\begin{figure}[t]
\centering
        \begin{subfigure}[b]{0.23\textwidth}
                \includegraphics[trim=1.3in 0.9in 1.3in 1in,clip,width=\textwidth]{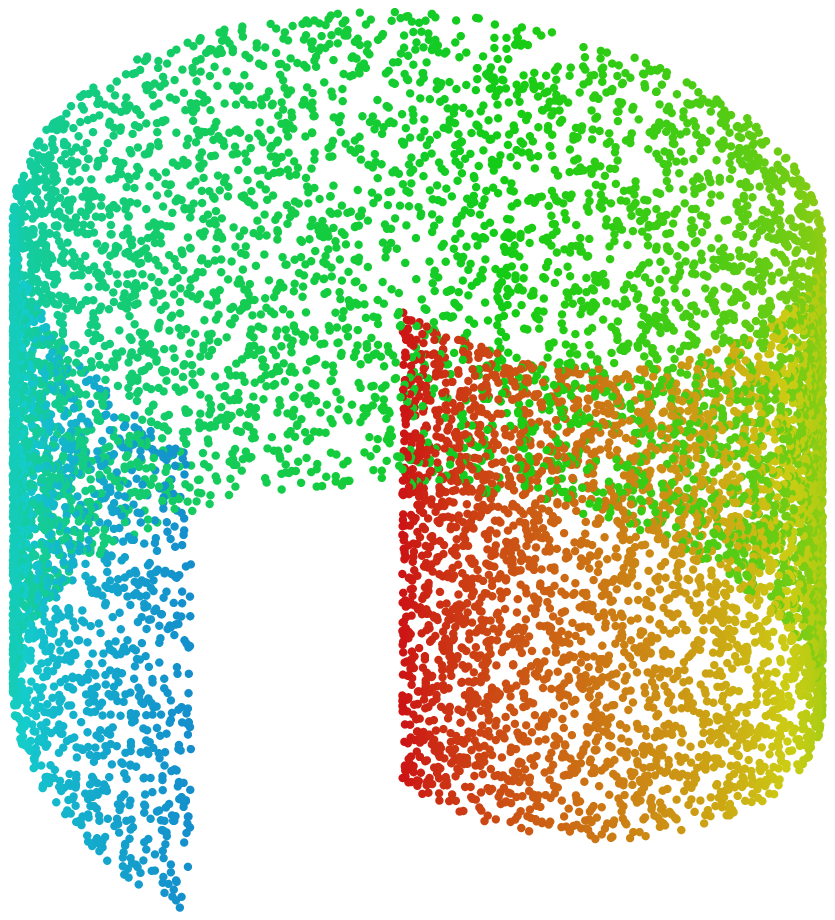}
        \end{subfigure}%
        ~
		\begin{subfigure}[b]{0.23\textwidth}
    		\includegraphics[trim=1.4in 0.9in 1.3in 1in,clip,width=\textwidth]{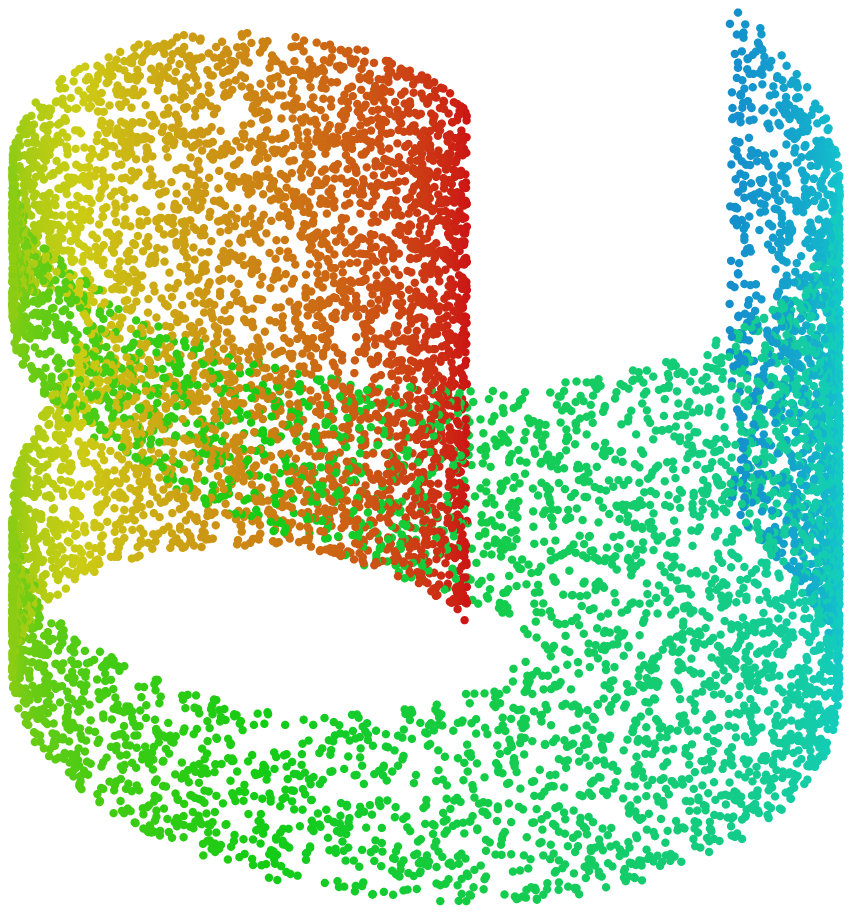}
	    \end{subfigure}%
        \\
        \begin{subfigure}[b]{0.23\textwidth}
        \begin{center}
                \includegraphics[trim=0.3in 0.1in 0.3in 0.5in,clip,width=0.07\textwidth]{arrow.eps}
        \end{center}
        \end{subfigure}%
        ~
        \begin{subfigure}[b]{0.23\textwidth}
        \begin{center}
                \includegraphics[trim=0.3in 0.1in 0.3in 0.5in,clip,width=0.07\textwidth]{arrow.eps}
        \end{center}
	    \end{subfigure}%
        \\
        \begin{subfigure}[b]{0.23\textwidth}
                \includegraphics[trim=0.8in 0.8in 0.8in 0.7in,clip,width=\textwidth]{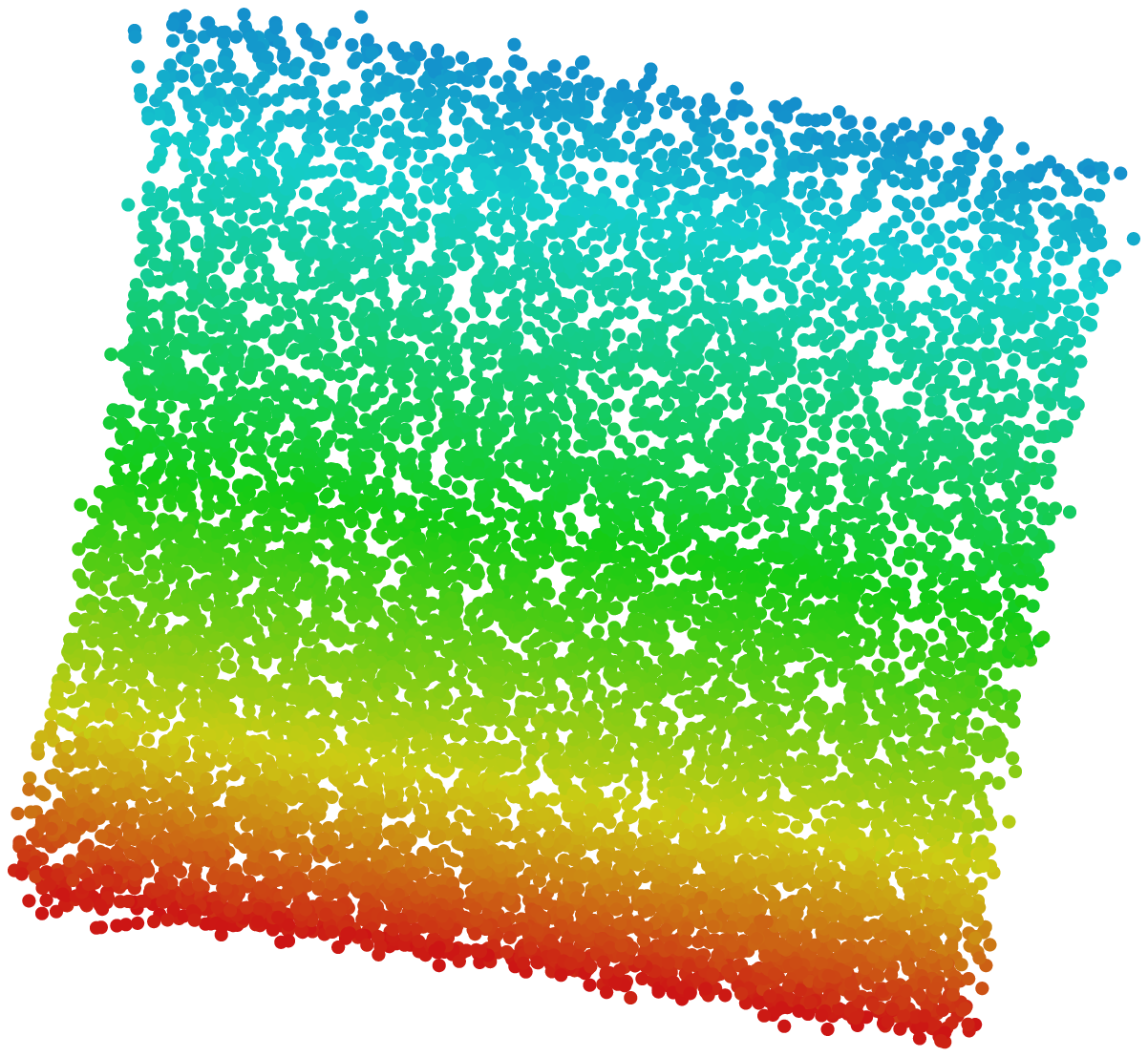}
                \caption{}
        \end{subfigure}%
        ~
        \begin{subfigure}[b]{0.23\textwidth}
                \includegraphics[trim=0.7in 0.8in 0.7in 0.4in,clip,width=\textwidth]{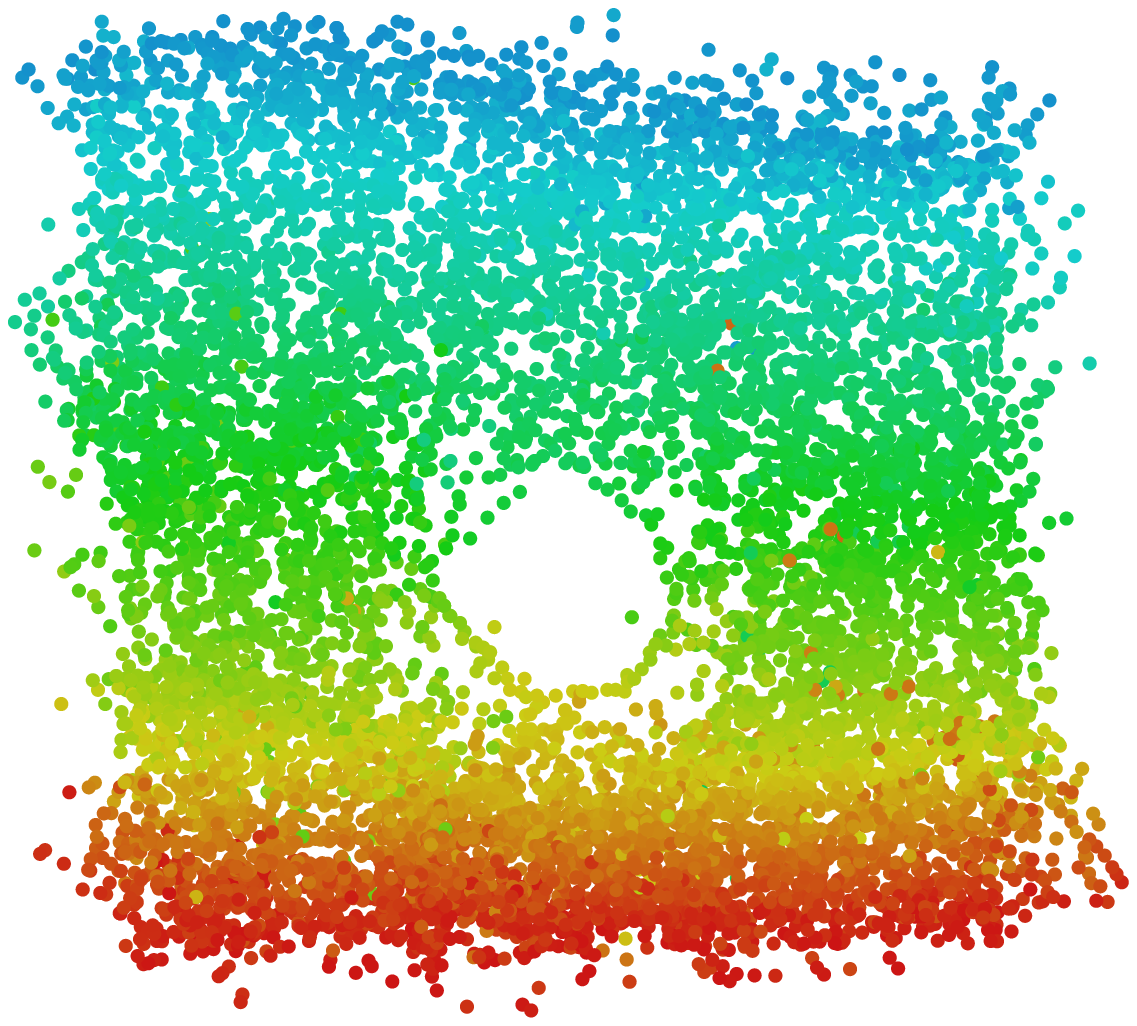}
                \caption{}
	    \end{subfigure}
        \vspace{2mm}
        \caption{{\small Applying Path-Based Isomap on (a) Swiss-Roll dataset with $N=10000$ and $P=846$, and (b) Swiss-Hole dataset with $N=10000$ and $P=930$.}}
        \label{fig_8_1}
\end{figure}

\subsection{Synthetic Datasets}
Swiss-Role is a typical dataset for testing manifold learning methods. Fig.\ref{fig_8_1}(a) shows the that Path-Based Isomap successfully unfolds a Swiss-Roll with $N=10000$ data points. It is notable that via the path-mapping scheme, degrees of freedom is dropped $82\%$. 

\begin{figure}[b]
\centering
        \begin{subfigure}[b]{0.35\textwidth}
                \includegraphics[trim=1in 0.6in 0.9in 0.4in,clip,width=\textwidth]{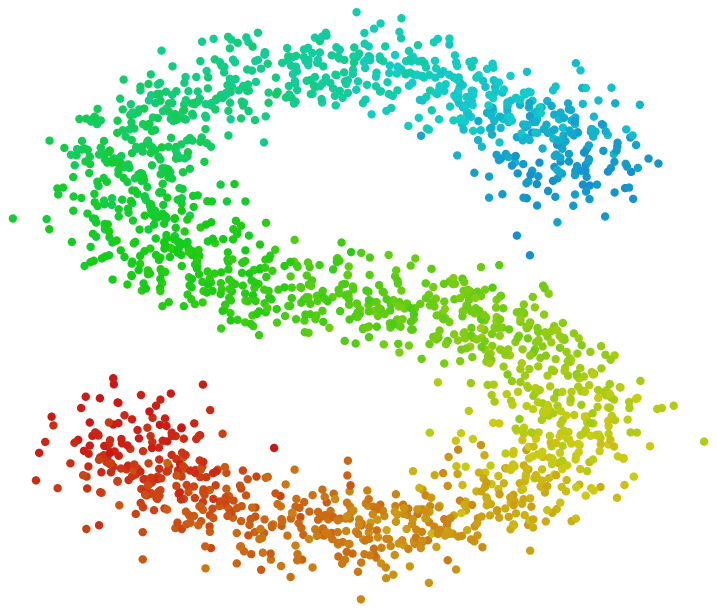}
        \end{subfigure}%
        \\ \vspace{1mm}
        \begin{subfigure}[b]{0.2\textwidth}
        \begin{center}
                \includegraphics[trim=0.3in 0.1in 0.3in 0.5in,clip,width=0.07\textwidth]{arrow.eps}
        \end{center}
        \end{subfigure}%
        \\%
        \begin{subfigure}[b]{0.35\textwidth}
                \includegraphics[trim=1.5in 0.2in 1.1in 0.2in,clip,width=\textwidth]{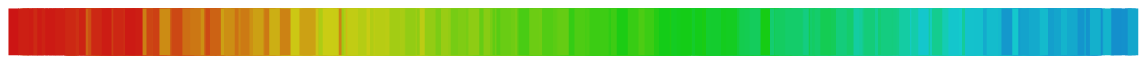}
        \end{subfigure}%
        \label{fig_Sshape}
        \vspace{5mm}
        \caption{{\small Investigating the effect of noise on Path-Based Isomap via a noisy S-shape one-dimensional manifold in $\mathcal{R}^2$. Number of data samples is $N=2000$ and the number of obtained paths is $185$.}}
\end{figure}

\begin{figure*}[t]    
\centering
\begin{tabular}{cc|ccc}
	\multicolumn{2}{c|}{\multirow{2}{*}{	
		\begin{minipage}[r]{0.3\textwidth}
			\vspace{-4mm}
			\includegraphics[trim=1.4in 1.1in 1.25in 1.2in,clip,width=0.8\textwidth]{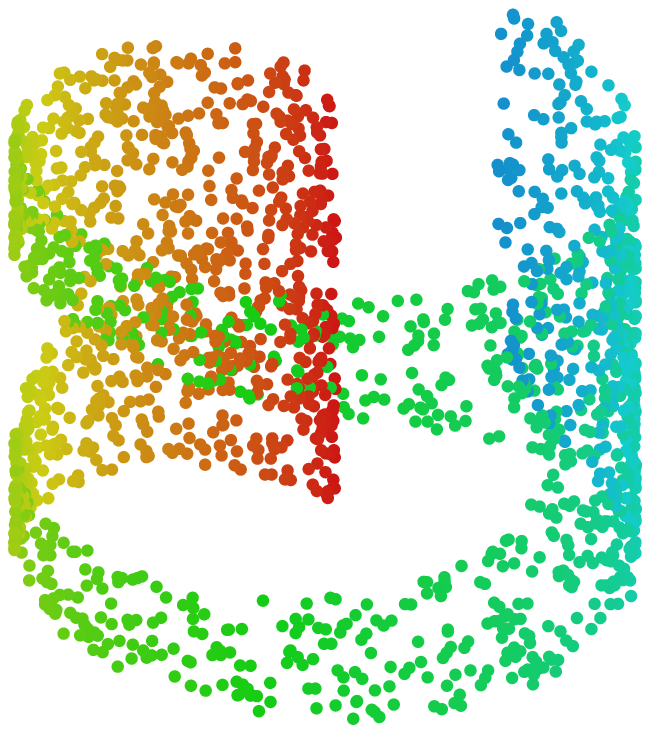}
		\end{minipage}    	    	    	
	}} & 
	\begin{subfigure}[c]{0.2\textwidth}
        \includegraphics[trim=0.7in 0.7in 0.7in 0.5in,clip,width=\textwidth]{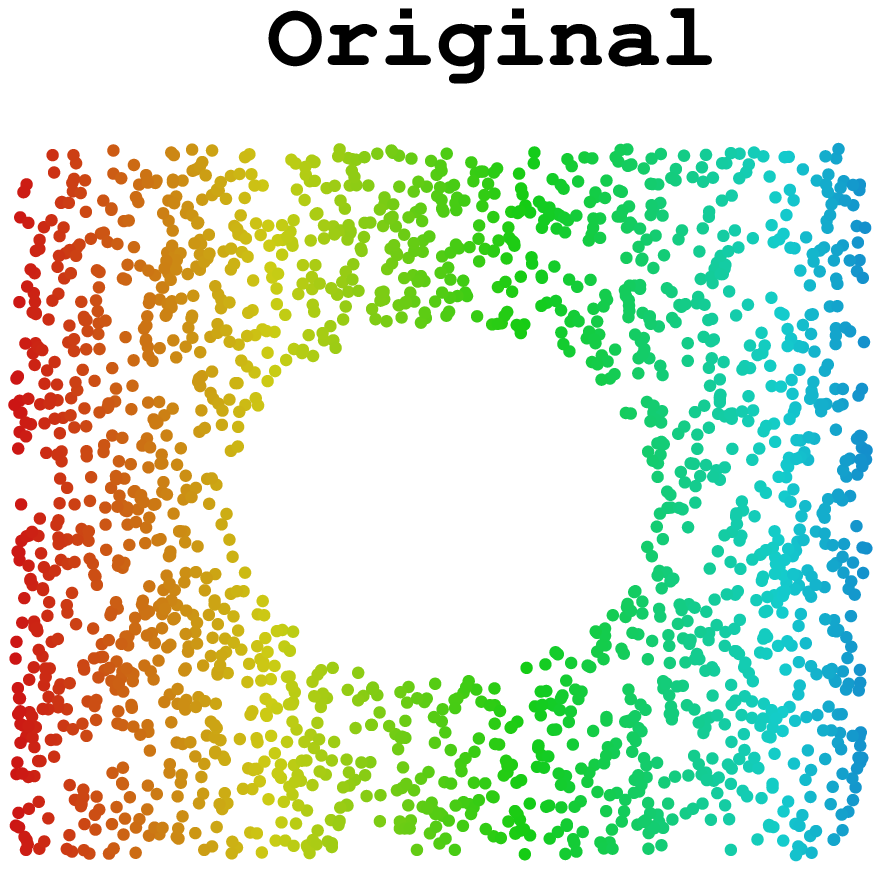}
    \end{subfigure} & 
	\begin{subfigure}[c]{0.2\textwidth}
    	\includegraphics[trim=0.7in 0.7in 0.4in 0.5in,clip,width=\textwidth]{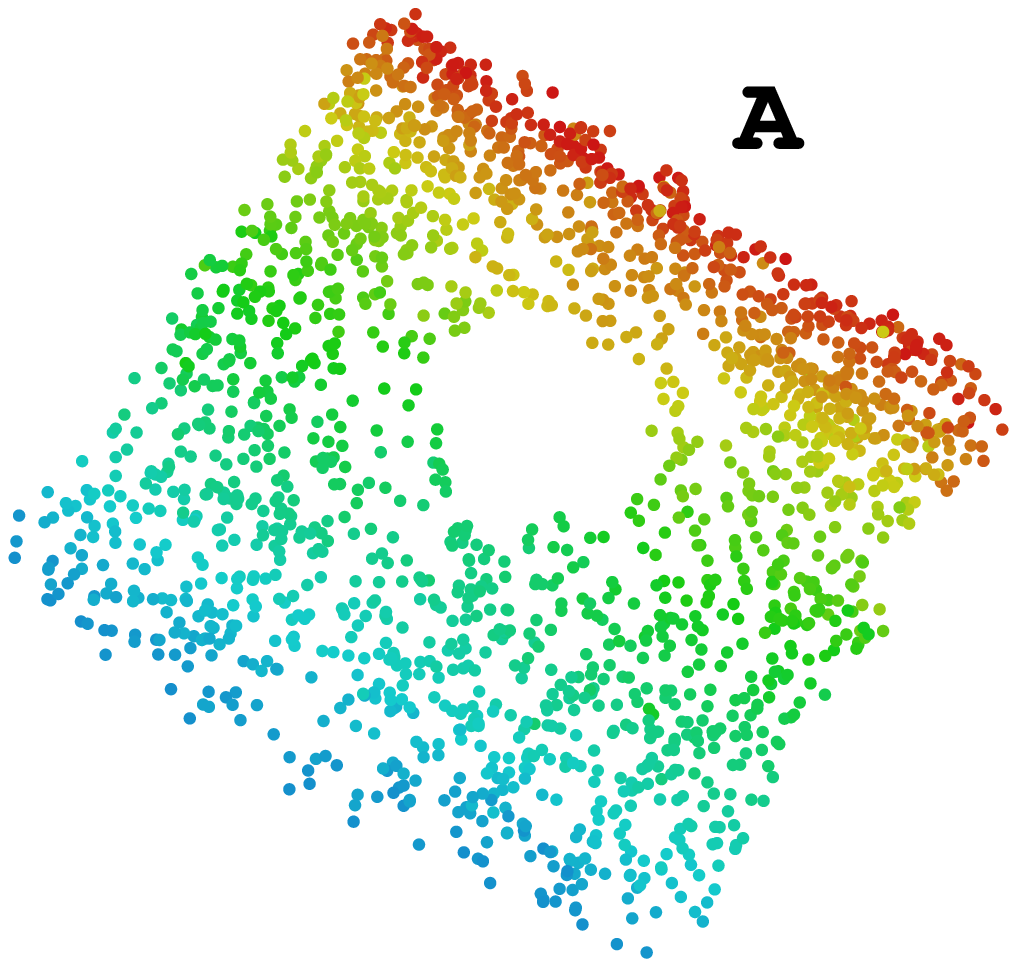}                
    \end{subfigure} &
	\begin{subfigure}[c]{0.2\textwidth}
        \includegraphics[trim=0.7in 0.7in 0.6in 0.5in,clip,width=\textwidth]{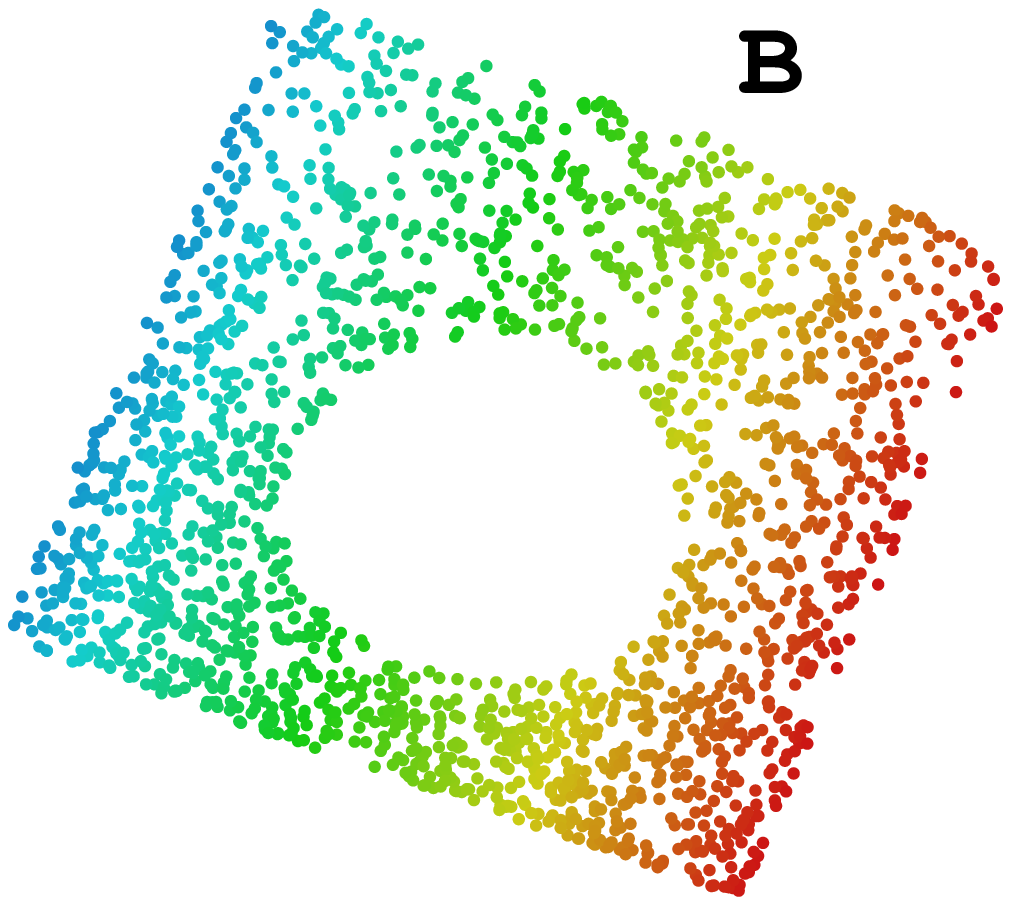}
    \end{subfigure}    
    \\ \multicolumn{2}{c|}{} & 
	\begin{subfigure}[c]{0.2\textwidth}
        \includegraphics[trim=0.7in 0.7in 0.7in 0.5in,clip,width=\textwidth]{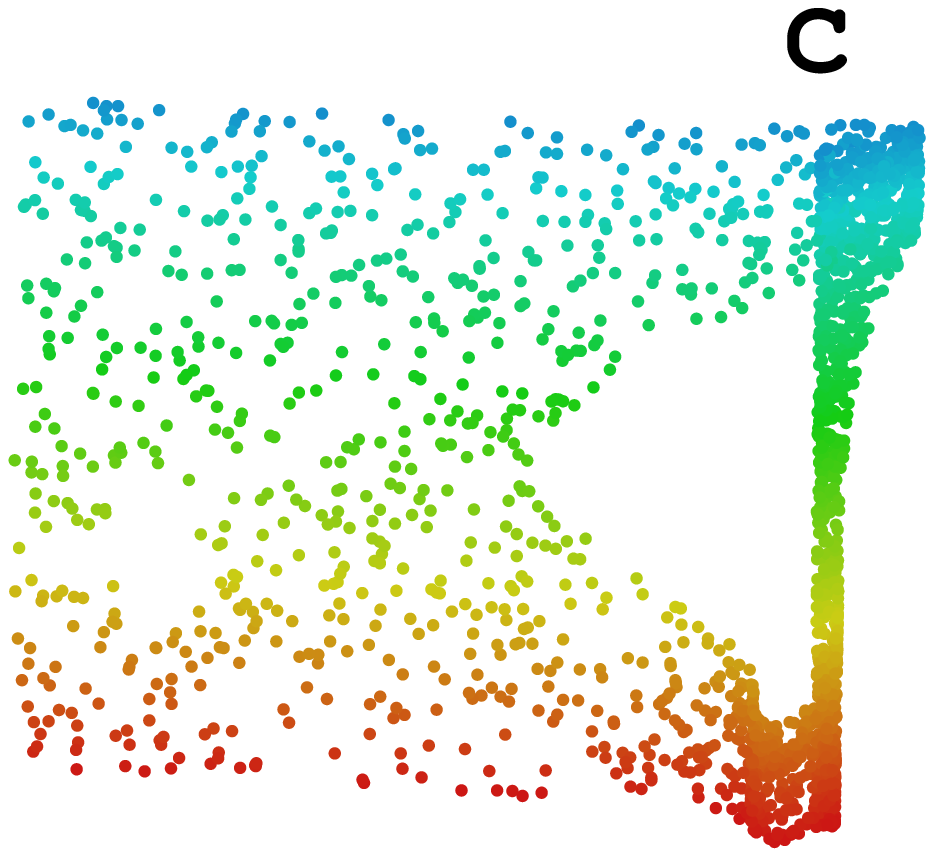}
    \end{subfigure} &
    \begin{subfigure}[c]{0.2\textwidth}
    	\includegraphics[trim=0.7in 0.7in 0.7in 0.5in,clip,width=\textwidth]{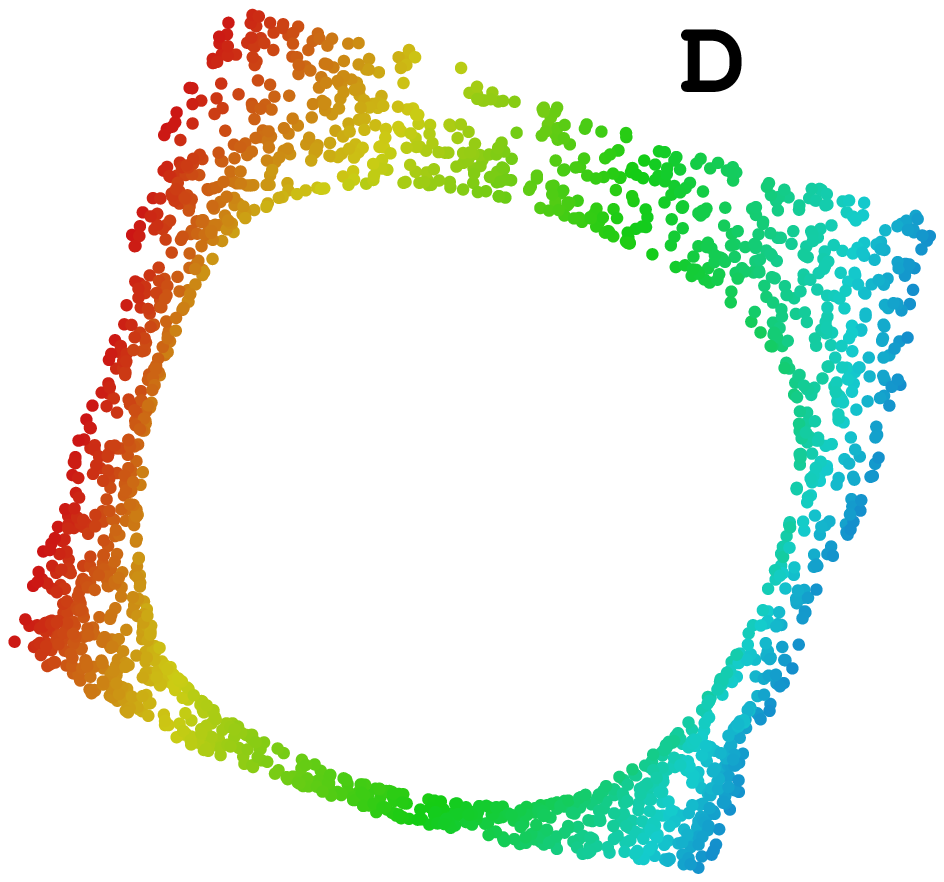}
    \end{subfigure} &
    \begin{subfigure}[c]{0.2\textwidth}
    	\includegraphics[trim=0.7in 0.7in 0.7in 0.5in,clip,width=\textwidth]{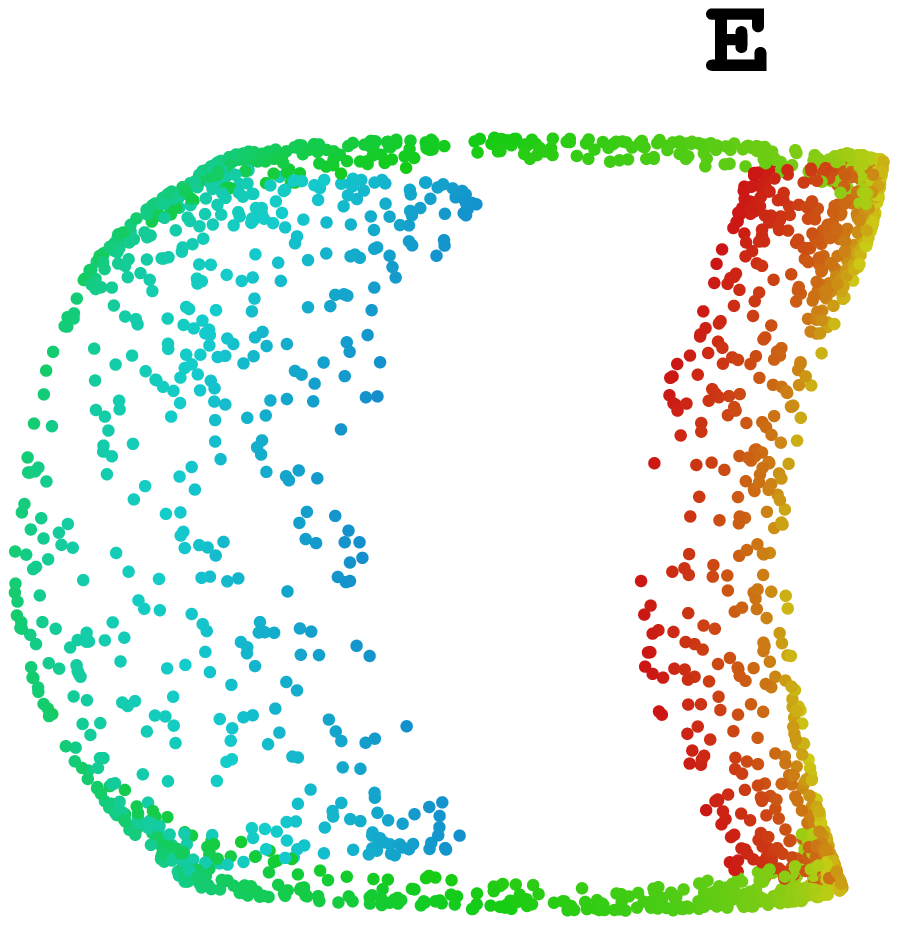}
    \end{subfigure}
\end{tabular}
	\vspace{3mm}
    \caption{{\small Comparison of performance for the proposed path-based method and four state-of-the-art algorithms on a Swiss-Hole dataset with $N=2000$. The result are obtained by Path-Based Isomap (A), Hessian LLE (B), LLE (C), Isomap (D) and Laplacian-Eigenmaps (E). Hessian LLE and Path-Based Isomap have outperformed other methods.}}
    \label{fig_perf_comp}
\end{figure*}

Another commonplace task for examining performance of an algorithm on non-convex geometrical structures is the Swiss-Hole dataset.
Fig.\ref{fig_8_1}(b) shows the performance of the Path-Based Isomap on a Swiss-Hole. The challenge, especially for Isomap, arises from the fact that for pairs on opposite sides of the hole, the shortest path on $G$ will no longer serve as the Euclidean distance in low-dimensional space. This might lead to the failure of the whole algorithm \cite{wang2008manifold}. However, Path-Based Isomap demonstrates acceptable resiliency to such non-normality. This effect can be understood as a consequence of (\ref{eq19}), in that good estimations in (\ref{eq19}) will correct poor ones to some extent. However, the correction causes the hole to be shrunk.

One of the main drawbacks of manifold learning methods involving shortest path calculation is their sensitivity to noise. Even one short-circuit may lead to miscalculation of many geodesic distances, and cause a drastic decline in performance. Thus the robustness of Path-Based Isomap was tested on a noisy S-shape dataset. The experiment demonstrates that the method is to an acceptable degree resistant to outliers. The averaging strategy might be again the reason behind this observation, since good estimations make for the poor ones.

\begin{figure}[t]
\centering
	\includegraphics[trim=0.2in 0.2in 0.2in 0.2in,clip,width=0.45\textwidth]{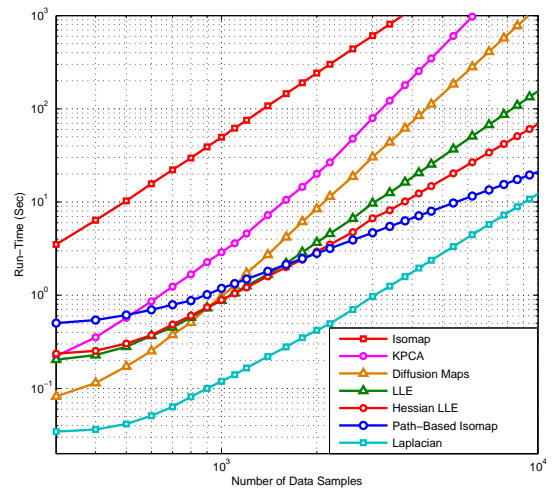}
	\caption{{\small Execution-time vs. number of data samples for $7$ manifold learning techniques including the proposed Path-Based approach. Experiments are done on a noisy S-shape manifold.}}
	\label{fig_runTime_comp}
\end{figure}

In Fig.\ref{fig_perf_comp} and Fig.\ref{fig_runTime_comp} we have illustrated a comparison among our proposed method and a number of rival methods for manifold learning. Fig.\ref{fig_perf_comp} demonstrates the performance of the proposed Path-Based Isomap method, compared to $4$ well-known rival methods. Experiments are done on a Swiss-Hole dataset consisting of $N=2000$ data samples which is known as a controversial dataset for most manifold learning techniques. We have utilized the DRtoolbox to implement the $4$ rival algorithms, which is known as an efficient and effective toolbox for manifold learning. In order to find the parameters for each algorithm in DRtoolbox, a precise grid-search is formed and the most appropriate parameters are chosen. As illustrated in Fig.\ref{fig_perf_comp}, Path-Based Isomap and Hessian LLE have been successful in unfolding the embedded manifold. In both methods, the hole has a small shift toward one of the sides. Hessian LLE has preserved the size of the hole, while in the path-based approach the hole is shrunk. Other methods such as Isomap, LLE and Laplacian-Eigenmaps have shown poor performances on this dataset. However, as can be seen, both Isomap and Path-Based Isomap have preserved the overal structure of data since both methods are considered as global approaches. Methods such as Laplacian-Eigenmaps and LLE are local methods and thus do not necessarily preserve the geometrical structure among far samples.

Fig.\ref{fig_runTime_comp} in this section demonstrates the time-complexity analysis of the proposed path-based approach compared to $6$ existing rival methods. Running-time of the methods on an S-shaped one-dimensional manifold are depicted as a function of number of data sample. Rival methods including Isomap, Kernel PCA, Diffusion Maps, LLE, Hessian LLE and Laplacian-Eigenmaps are again implemented via DRtoolbox. The methods are chosen so they have been claimed to have high efficiency or appropriate accuracy. From Fig.\ref{fig_runTime_comp} it is evident that the proposed path-based method have a considerable lower slope in a log-log plot, meaning that the method will outperform all the rival algorithms for sufficiently large $N$. Up to $N=10000$ the proposed method has already surpassed $5$ rival algorithms. Observations agree to theoretical analysis in Section $7$.

\subsection{Real-World Datasets}
A canonical problem in dimensionality reduction is pose estimation. Fig.\ref{fig_8_1} illustrates statue-face database, consisting of $698$ $64\times 64=4096$ pixel images, rendered with different camera angles and random light directions \cite{statueFace}.  
Data samples are believed to lie on a smooth manifold in $\mathcal{R}^{4096}$ \cite{tenenbaum2000global}. We apply Path-Based Isomap to discover the compact representation for the dataset. Interestingly, the algorithm unfolds the $2$-dimensional manifold of the original $4096$-dimensional data samples in $\mathcal{R}^2$. The horizontal and vertical axis are tightly related to horizontal and vertical angles of the camera. 
To plot this figure we have extracted the first $240$ linear components via PCA prior to applying Path-Based Isomap. This preprocessing improved the results, meaning that there is also a significant linear redundancy among data points. 

\begin{figure}[b]
\centering
    \includegraphics[trim=0.5in 0.3in 0.5in 0.3in,clip,width=0.45\textwidth]{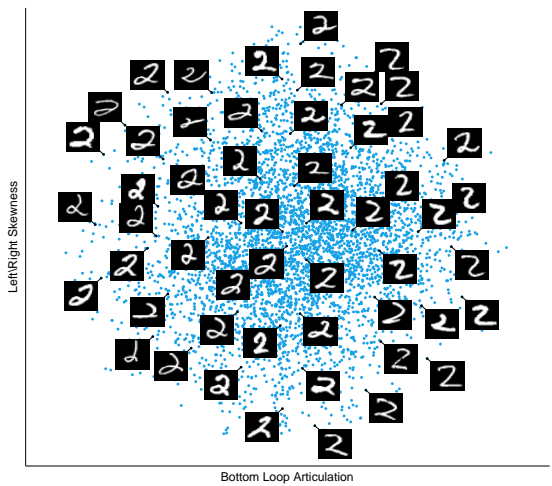}
    \caption{{\small The result of applying Path-Based Isomap on MNIST database. There are $5000$ images of handwritten `$2$'s images in the database. There is clearly a meaningful relation between place of data points and geometrical features of their corresponding images.}}
    \label{fig_8_4}
\end{figure}

\begin{figure}[t]
\centering
    \includegraphics[trim=0.45in 0.3in 0.5in 0.3in,clip,width=0.45\textwidth]{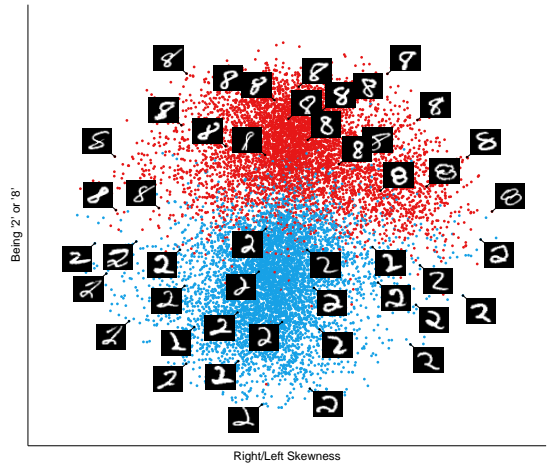}
    \caption{{\small Applying Path-Based Isomap on $11800$ images of `$2$'s and `$8$'s in MNIST database to obtain $2$-dimensional representations. Two clusters are formed that mainly contain one type of digit. Besides, there is a link between horizontal placing and left/right skewness in the images.}}
    \label{fig_8_5}
\end{figure}

Fig.\ref{fig_8_4} and Fig.\ref{fig_8_5} illustrate the performance of Path-Based Isomap on MNIST \cite{MNIST}. MNIST is a well known image classification database of handwritten numbers. Despite the fact that these images do not necessarily lie on a manifold, experiments reveal that the proposed method achieves a good performance on them. In Fig.\ref{fig_8_4} the method is applied on handwritten `$2$' images to discover their compact description in $\mathcal{R}^2$. As expected, the samples are placed according to articulation of the bottom loop and horizontal skewness of the structure. In Fig.\ref{fig_8_5} the method is applied to the combined datasets of handwritten images of `$2$'s and `$8$'s. It is evident that different digits are largely separated along the Y-axis. Moreover, `$2$'s and `$8$'s are ordered according to horizontal skewness along the X-axis. It is also notable that `$2$'s on the left side of the plane have stronger bottom loop articulations.

\begin{figure}[t]
\centering
    \includegraphics[trim=0.5in 0.3in 0.5in 0.3in,clip,width=0.45\textwidth]{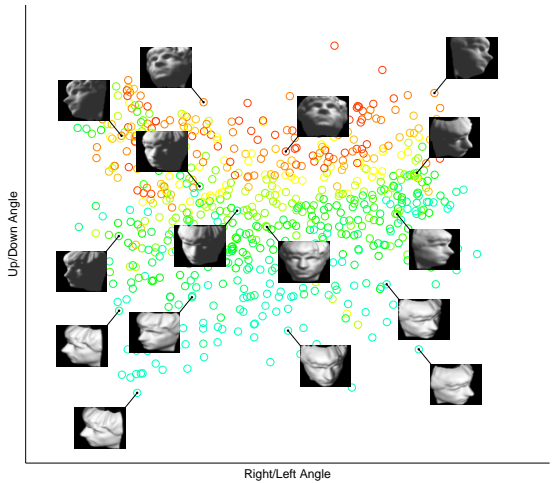}
    \caption{{\small Applying Path-Based Isomap on $64\times 64$ gray-scale images of the statue-face database. Some images are shown to illustrate performance of algorithm.}}
    \label{fig_8_3}
\end{figure}

\section{Conclusion and Future Work}
In this paper we proposed Path-Based Isomap, a new efficient method for dimensionality reduction. The method exploits shortest paths instead of samples to compute the compact representation faster. Path mapping and path selection schemes were also developed to preserve geodesic distances in the representation space. The fewer number of paths leads to significant cost savings in our approach.

It was shown throughout the paper that most virtues of LLE, Isomap and Laplacian Eigenmaps are shared by Path-Based Isomap. Experiments showed the method works surprisingly well on non-convex manifolds, as well as real-world databases like MNIST and face-statue. Moreover, it was shown that the method is to some extent resilient against noise. 

The most encouraging achievement of this paper is that the method works remarkably faster than most of rival methods, especially on large scale datasets. This improvement was confirmed by our theoretical analysis of its memory and time complexity. Since dimensionality reduction methods are now very popular in image classification and pattern recognition tasks, Path-Based Isomap should find widespread use in applications. 

Since the number of found paths has a direct effect on complexity of the method, and regarding suboptimal performance of SSPC, there is considerable space for improvement. Besides , there are still two shortcomings shared between rival approaches and ours. First, the method is sensitive to outliers causing short-circuits, because they could mislead Dijkstra's algorithm in the first stage. Second, we assume the sampling density to be nearly uniform. Further empirical and theoretical research could shed some light on these issues.


%

\appendices
\section{Network Rigidity}
The method discussed before for the shortest path covering of the data samples guarantees that each sample lies on at least one path (line). However, the uniqueness of low-dimensional representation under these circumstances must be further investigated. In other words many radically different representations of a particular set of data samples may result in one set of shortest paths, which implies that loss of information is possible. 

Fig. \ref{fig_6_1_a} shows a set of $28$ data samples spread in a two-dimensional space which are covered with $3$ separated paths. It can be easily observed that the obtained network of paths is not rigid since each path is free to move independently with respect to others. The constraints of placing all the data samples in their corresponding lines with their corresponding orders and distances are not enough to obtain a unique description of all the data sample positions. It is clear that without any intersection there is no objective function and hence no unique solution.

As shown in Fig. \ref{fig_6_1_b} adding three more paths to the network results into a rigid structure that uniquely describes the proximity information of the data samples. For a rigid structure, there is no degree of freedom except rotation and translation of the whole object in $\mathcal{R}^K$ that do not affect neighborhood information. The translation can be mathematically modeled as a constant $K$-dimensional vector added to all $\boldsymbol{\xi}_p$s. The rotation is also modeled with a $K\times K$ unitary matrix affecting all $\boldsymbol{\xi}_p$ and $\hat{\boldsymbol{v}}_p$ vectors. Therefore we can express the following definition of a rigid network:

\begin{figure}[t]
\centering
        \begin{subfigure}[b]{0.22\textwidth}
                \includegraphics[trim=0.8in 0.3in 0.8in 0.3in,width=\textwidth]{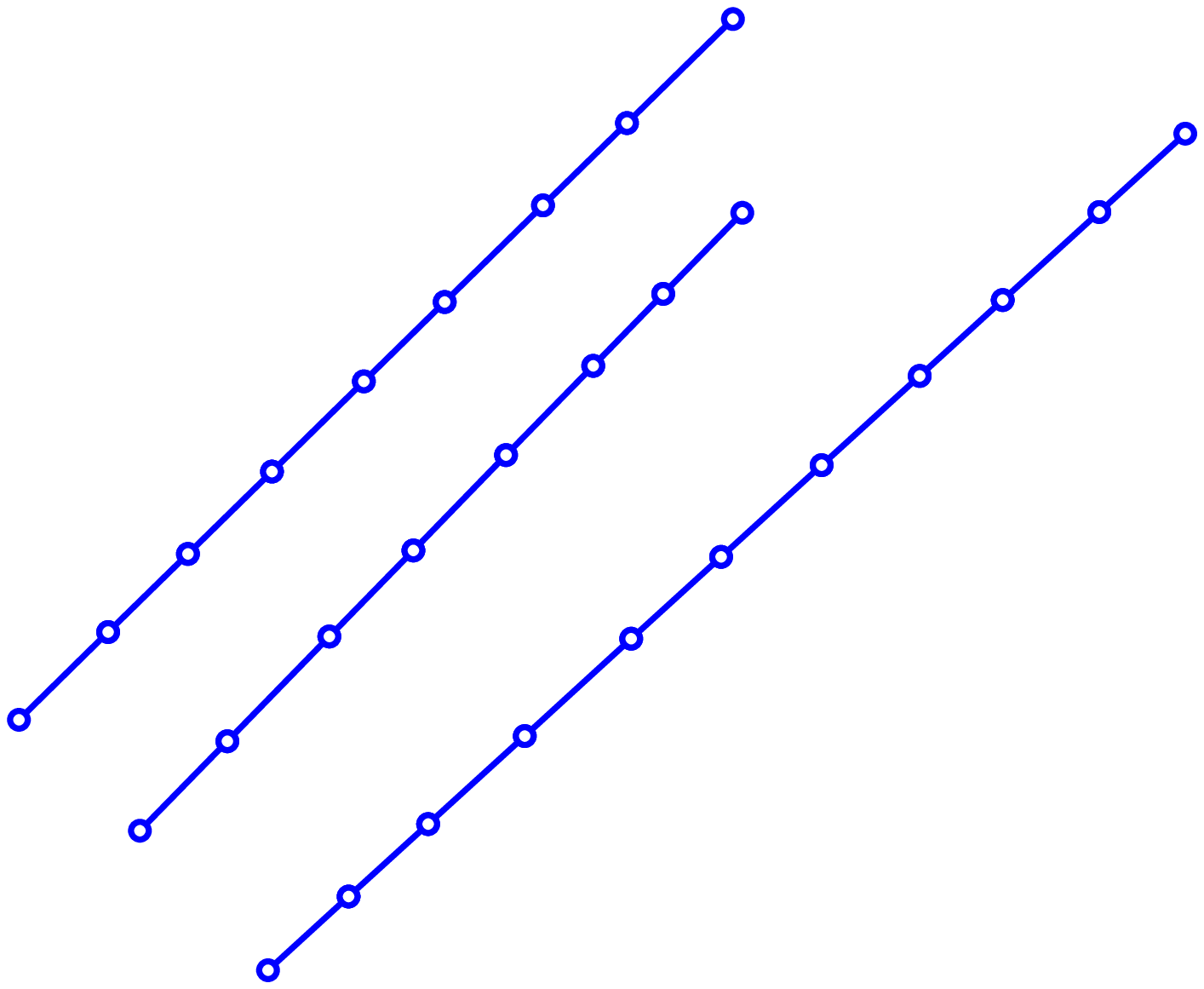}
                \caption{}
                \label{fig_6_1_a}
        \end{subfigure}%
        \begin{subfigure}[b]{0.22\textwidth}
                \includegraphics[trim=0.8in 0.3in 0.8in 0.3in,width=\textwidth]{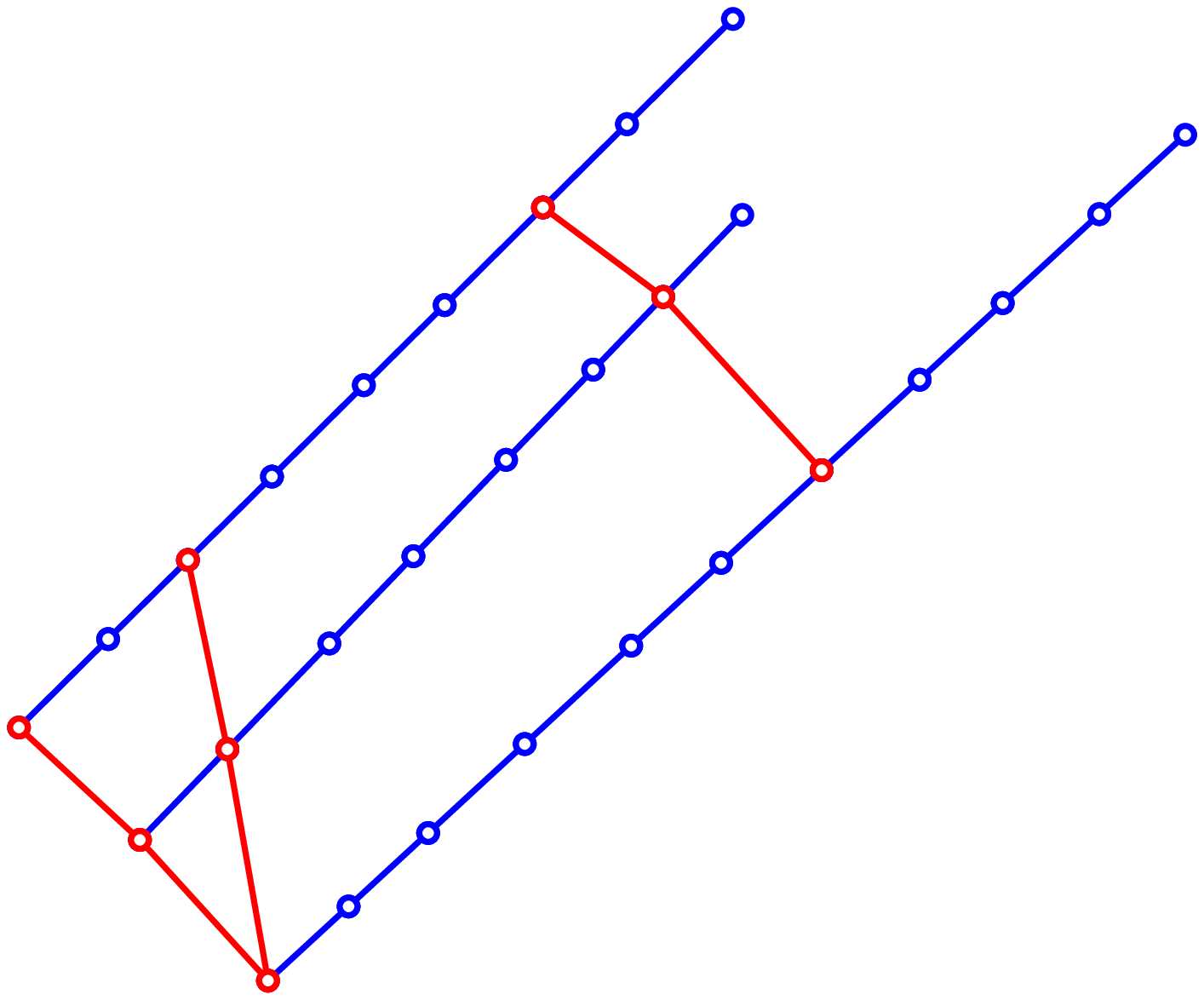}
                \caption{}
                \label{fig_6_1_b}
        \end{subfigure}
        \caption{{\small An example of a non-rigid and a rigiditized network of intersecting lines. (a) $28$ two-dimensional data samples are covered using $3$ isolated lines. Although all the data samples are covered, the resulting network in not rigid. (b) Adding $3$ intersecting lines as shown (red lines) to the network results into a rigid structure.}}
 \label{fig_6_1}
\end{figure}

\begin{framed}
\begin{myDefinition}
A rigid network of intersecting paths is defined as a set of lines in $\mathcal{R}^K$ whose data samples can be uniquely and unambiguously mapped into the representation space except for a total translation by a constant vector and a rotation by any unitary matrix.
\end{myDefinition}
\end{framed}

In order to reach a rigid network a compensation strategy must be utilized after the shortest path covering stage. So sufficient paths to stabilize the current structure could be found and added to the set $\Omega$. It has been experimentally observed that the stochastic shortest path covering method introduced in section 4 of the submitted paper, usually results directly in a rigid network and does not need this step. The reason behind this phenomenon is the numerous intersections obtained by the stochastic path covering algorithm. However, few stabilizing lines should be added in some cases specially when the number of data samples is not large enough. The proposed strategy for the stabilization of a network is explained as follows.

Assume a set of data samples in high-dimensional space are covered by a set of $P$ shortest paths denoted by $\Omega=\left\{\mathcal{L}_1,\mathcal{L}_2,...,\mathcal{L}_P\right\}$. Also assume that the lines in $\Omega$ are intersected in a set of nodes denoted by $\tilde{V}=\left\{\tilde{v}_{i,j}\right\},\quad i,j\in \left\{1,2,...,P\right\}$ where $\tilde{v}_{i,j}$ denotes the common node between the $i$th and the $j$th lines. If lines $\mathcal{L}_i$ and $\mathcal{L}_j$ do not cross each other then $\tilde{v}_{i,j}$ does not exist.

Many sub-networks of a rigid network are also rigid. Formally speaking, a rigid sub-network is a set of lines, denoted as $\mathcal{S}=\left\{\mathcal{L}_{\zeta_1^{\mathcal{S}}},\mathcal{L}_{\zeta_2^{\mathcal{S}}},...,\mathcal{L}_{\zeta_{L_\mathcal{S}}^{\mathcal{S}}}\right\}$,  which are shown to be rigid with respect to each other. Here if $L_\mathcal{S}$ is the number of lines in $\mathcal{S}$, $\zeta_i^{\mathcal{S}}\in\left\{1,2,...,P\right\}$ for $i=1,2,...,L_\mathcal{S}$ are the indices of $L_\mathcal{S}$ lines (paths) in $\Omega$.

\begin{framed}
\begin{myTheorem}
The smallest possible rigid sub-network with non-zero volume in $\mathcal{R}^K$ is a geometrical structure consisting of $\frac{1}{2}K(K+1)$ lines called a hyper-pyramid. A hyper-pyramid $\Delta= \left\{\mathcal{L}_{\zeta_1^{(\Delta)}},\mathcal{L}_{\zeta_2^{(\Delta)}},...,\mathcal{L}_{\zeta_{\frac{1}{2}K(K+1)}^{(\Delta)}}\right\}$ can be identifies by the following properties.

\vspace{3mm}
For each feasible $\mathcal{S}$, we should look for $K+1$ subsets of $\mathcal{S}$, denoted by $\mathcal{D}_i$s ($i=1,2,...,K+1$), so that each subset represents $K$ different lines of the object. Each two distinct $\mathcal{D}_i$s must share exactly one line:
\begin{center}
$\exists\hspace{1mm}\mathcal{D}_1,\mathcal{D}_2,...,\mathcal{D}_{K+1}\subset \left\{\zeta_1^{(\Delta)},\zeta_2^{(\Delta)},...,\zeta_{\frac{1}{2}K(K+1)}^{(\Delta)}\right\}$\\
\vspace{3mm}
$\Rightarrow 
\left\{
\begin{array}{l l}
|\mathcal{D}_m|=K \\
|\mathcal{D}_m\cap\mathcal{D}_n|=1
\end{array} \right.\
\quad m\neq n, \quad  m,n\in\left\{1,2,...,K+1\right\}$
\end{center}
Also the lines in each subset, $\mathcal{D}_i$, must cross each other in at least one common data sample:
\begin{center}
$\mathcal{D}_i=\left\{d_1^{(i)},d_2^{(i)},...,d_K^{(i)}\right\}$\\
$(\tilde{v}_{d_1^{(i)},d_2^{(i)}}=\tilde{v}_{d_1^{(i)},d_3^{(i)}}=...=\tilde{v}_{d_K^{(i)},d_{K+1}^{(i)}})\in \tilde{V}$\\ \vspace{1mm}
$i=1,2,...,K+1$
\end{center}
\vspace{3mm}
The only exception is the special case of $K=1$, where the above conditions should be replaced by simply existence of a single line.
\end{myTheorem}
\end{framed}

The $|.|$ operator in the above expressions denotes the number of members in a set. For the simple cases of $K=1,2$ and $3$ the corresponding hyper-pyramid is a single line, two-dimensional triangle and three-dimensional pyramid respectively.

The proof for the Theorem 1 is given in Appendix B. The importance of the Theorem 1 is providing a systematic procedure to find many initial rigid sub-networks (hyper-pyramids) within a network of intersecting lines. It is observed experimentally that running the shortest path covering algorithm on a densely sampled manifold results in several hyper-pyramids.
	
\begin{framed}
\begin{myTheorem}
If a single line $\mathcal{L}_i$ has two intersections with a rigid sub-network $\mathcal{S}$, adding $\mathcal{L}_i$ to the $\mathcal{S}$ results in a new rigid sub-network. This can be expressed as:
\begin{center}
$\exists j,k\in \left\{1,2,...,L_{\mathcal{S}}\right\}$\\
\vspace{1mm}
$\Rightarrow \left\{\tilde{v}_{i,\zeta_j^{(\mathcal{S})}},\tilde{v}_{i,\zeta_k^{(\mathcal{S})}}\right\}\subset\tilde{V}$
\end{center}
\end{myTheorem}
\end{framed}

Proof of Theorem 2 can be found in Appendix C. Theorem 2 enables us to gradually build a rigid network by offering an approach to merge a single line and a rigid network, and obtain a larger rigid structure if they have at least two intersections. These rigidity checking and merging procedures can be iterated until the achieved rigid network cannot grow larger anymore. We can simply start the procedure with a hyper-pyramid. It is not possible to start by single lines since any attached line will be in almost the same direction, leaving no option to grow in other dimensions. In fact, as discussed in Theorem 1, the initial rigid sub-networks should have non-zero volume in $\mathcal{R}^K$.

If adding lines ended up in a sub-network that can not grow larger and does not include all the samples, we should add further paths to insure rigidity. Hereby, we could determine that low-dimensional representation will be unique. However, this usually does not happen, that means usually no further paths are needed in practice.

\section{On Rigidity of Hyper-pyramids}
In this section a proof for the Theorem 1 is derived. Theorem 1 states that a hyper-pyramid is always a rigid sub-network.

According to the definition of a hyper-pyramid in Theorem 1, such sub-networks contain $K+1$ corner points in $\mathcal{R}^K$ space. Each corner point is the intersection of $K$ different lines (for example a triangle has $3$ corner points while a pyramids contains $4$). From now on, for a $K$-dimensional hyper-pyramid we denote the mentioned corner points as $\boldsymbol{r}_1,\boldsymbol{r}_2,...,\boldsymbol{r}_{K+1}$. In order to cancel the effect of total translation in the representation, we assume that a constant vector $T=\frac{1}{K+1}\left(\boldsymbol{r}_1+\boldsymbol{r}_2+⋯+\boldsymbol{r}_{K+1}\right)$ is subtracted from all the corner points to place the center of gravity on the origin. Therefore we would have the following condition:
\begin{equation}
\label{eq20}
\sum_{k=1}^{K+1}\boldsymbol{r}_k=0
\end{equation}

Corner points of a hyper-pyramid are completely interconnected, i.e. there is a line (path) between each two corners. This indicates that the distances among all pairs of corner points are assumed to be known, which leads to the following system of quadratic equations:
\begin{equation}
\label{eq21}
\left\{
  \begin{array}{l l}
    ||\boldsymbol{r}_2-\boldsymbol{r}_1||_2^2=D_1^2 \\    
    ||\boldsymbol{r}_3-\boldsymbol{r}_1||_2^2=D_2^2 \\
    \vdots \\
    ||\boldsymbol{r}_{K+1}-\boldsymbol{r}_K||_2^2=D_{\frac{1}{2}K(K+1)}^2
  \end{array} \right.\
\end{equation}
As discussed before, the geodesic distances, $D_i$s ($i=1,2,...,\frac{1}{2}K(K+1)$), are known as a result of the shortest path covering stage.

In order to prove the rigidity of hyper-pyramid structures, we will show that the set of solutions satisfying the above system of equations, differ only in an arbitrary rotation and translation.

Let us define the vector $X$ as follows:
\begin{gather}
\label{eq22}
\begin{array}{l l}
X_{\alpha_{i,i}}=||\boldsymbol{r}_i||_2^2 \\
X_{\alpha_{i,j}}=\boldsymbol{r}_i^{T}\boldsymbol{r}_j
\end{array}
\\
\alpha_{i,j}\in\left\{1,2,...,\frac{1}{2}(K+1)(K+2)\right\}, \quad {i,j=1,2,...,N, \atop i<j}
\end{gather}
where the parameterized indices $\alpha_{i,j}$ are solely used to facilitate the construction of $X$, and can be defined in any arbitrary order. Left sides of expressions in the equations of (\ref{eq21}) can be rewritten using the components of $X$, since there is the following relation:
\begin{equation}
\label{eq23}
||\boldsymbol{r}_i-\boldsymbol{r}_j||_2^2=X_{\alpha_{i,i}}+X_{\alpha_{j,j}}-2X_{\alpha_{i,j}}=D^2_{\beta_{i,j}}
\end{equation}

Again, $\beta_{i,j}$ indices are solely defined to facilitate the ordering of equations.

In addition, inner product of each corner point, $\boldsymbol{r}_1,\boldsymbol{r}_2,...,\boldsymbol{r}_{K+1}$, by both sides of (\ref{eq20}) gives an extra $K+1$ linear equations. The collection of all the mentioned linear dependencies results in the following $\frac{1}{2}(K+1)(K+2)\times \frac{1}{2}(K+1)(K+2)$ determined system of linear equations:
\begin{equation}
\label{eq24}
\left\{
  \begin{array}{l l}
    X_{\alpha_{i,i}}+X_{\alpha_{j,j}}-2X_{\alpha_{i,j}}=D^2_{\beta_{i,j}} \quad i,j=1,2,...,K+1\\           
    \sum_{j=1}^{K+1}X_{\alpha_{i,j}}=0 \quad i=1,2,...,K+1,\quad i>j
  \end{array} \right.\    
\end{equation}

The matrix formulation of above equations can be written as follows:
\begin{equation}
\label{eq25}
AX=D
\end{equation}
where $A$ is a known $\frac{1}{2}(K+1)(K+2)\times \frac{1}{2}(K+1)(K+2)$ matrix. It can be shown that the matrix $A$ in (\ref{eq25}) is invertible and the components of $X$ which represent all possible inner products of corner points, can be uniquely determined. In (\ref{eq26}) the $A$ matrix for the case of $K=2$ has been shown.
\begin{equation}
\label{eq26}
A=\left[
\begin{array}{cccccc}
1&1&0&-2&0&0\\
1&0&1&0&-2&0\\
0&1&1&0&0&-2\\
1&0&0&1&1&0\\
0&1&0&1&0&1\\
0&0&1&0&1&1
\end{array}
\right]
\end{equation}

Invertibility of $A$ states that the inner products among all pairs of the corner points are uniquely determined by the set of non-linear equations in (\ref{eq21}). In order to finalize the proof we must show that any two structures that have the same inner products among their corresponding corner points would differ only in a rigid transformation. For this purpose, assume that there are two different sets of solutions for the equations in (\ref{eq21}), denoted by $\boldsymbol{r}_i$ and $\tilde{\boldsymbol{r}}_i$, $i=1,2,...,K+1$. Based on the previous discussions, we have:
\begin{gather}
\label{eq27}
\boldsymbol{r}_i^T\boldsymbol{r}_j=\tilde{\boldsymbol{r}}_i^T\tilde{\boldsymbol{r}}_j \\
i,j=1,2,...,K+1 \nonumber
\end{gather}

Let us decompose each vector $\boldsymbol{r}_i$ into a non-negative value $r_i$, and a unit length vector $\hat{\boldsymbol{r}}_i$, which represent the length and the direction of $\boldsymbol{r}_i$ respectively.
\begin{gather}
\label{eq28}
\boldsymbol{r}_i=r_i\hat{\boldsymbol{r}}_{i}\\
||\hat{\boldsymbol{r}}_i||_2=1 \nonumber
\end{gather}

According to (\ref{eq27}), $r_i=\tilde{r}_i$ for all $i\in\left\{1,2,...,K+1\right\}$, so (\ref{eq27}) reduces to the following set of equalities:
\begin{gather}
\label{eq29}
\hat{\boldsymbol{r}}_i^T\hat{\boldsymbol{r}}_j=\tilde{\hat{{\boldsymbol{r}}}}_i^T\tilde{\hat{\boldsymbol{r}}}_j \\
i,j=1,2,...,K+1 \nonumber
\end{gather}

Both $\hat{\boldsymbol{r}}_i$ and $\tilde{\hat{{\boldsymbol{r}}}}_i$ are unit length vectors in $\mathcal{R}^K$, and thus are related through a particular rotation transform:
\begin{equation}
\label{eq30}
\tilde{\hat{{\boldsymbol{r}}}}_i=U_i\hat{\boldsymbol{r}}_i
\end{equation}
where $U_i$ is a $K\times K$ unitary matrix mapping $\hat{\boldsymbol{r}}_i$ onto $\tilde{\hat{\boldsymbol{r}}}_i$. Therefore we have:
\begin{gather}
\label{eq31}
\hat{\boldsymbol{r}}_i^T\hat{\boldsymbol{r}}_j=\hat{\boldsymbol{r}}_i^{T}U_i^{T}U_j\hat{\boldsymbol{r}}_j \\
\Rightarrow 
\hat{\boldsymbol{r}}_i^{T}\left(U_i^{T}U_j-I\right)\hat{\boldsymbol{r}}_j=0 \nonumber \\
\nonumber \\
i,j=1,2,...,K+1 \nonumber
\end{gather}

Since (\ref{eq31}) holds for all the unitary vectors associated with any arbitrary hyper-pyramid in $\mathcal{R}^K$, we may conclude that $U_i^{T}U_j=I$ for $\forall i,j\in\left\{1,2,...,K+1\right\}$. And finally:
\begin{gather}
\label{eq32}
U_1=U_2=...=U_{K+1}=U \\
\boldsymbol{r}_i=U\tilde{\boldsymbol{r}}_i \nonumber \\
\nonumber \\
i=1,2,...,K+1 \nonumber
\end{gather}

It should be reminded that the translation has been already canceled by shifting the center of gravity for each set of solutions to the origin. Therefore all possible sets of solution for a hyper-pyramid in $\mathcal{R}^K$ only differ in a translation and rotation which do not affect the proximity information of data samples.

\section{Merging Rigid Subnetworks}
This section provides the proof for Theorem 2. Based on Theorem 2, any line that has at least two intersections with a rigid sub-network, is totally rigid with respect to the sub-network. 

Assume a line $\mathcal{L}$ with starting point $\boldsymbol{\xi}$ and direction vector $\hat{\boldsymbol{v}}$ has two data samples, namely $\boldsymbol{r}_1$ and $\boldsymbol{r}_2$, shared with a rigid sub-network $\mathcal{S}$. The associated distances from the starting point of the line to the shared samples are denoted by $L_1$ and $L_2$ respectively. Hence we have:
\begin{equation}
\label{eq33}
\left\{
  \begin{array}{l l}
    \boldsymbol{r}_1=\boldsymbol{\xi}+L_1\hat{\boldsymbol{v}} \\    
    \boldsymbol{r}_2=\boldsymbol{\xi}+L_2\hat{\boldsymbol{v}}
  \end{array} \right.\
\end{equation}
	
Using simple algebra, following relations are obtained for the parameters of $\mathcal{L}$ in terms of shared data samples $\boldsymbol{r}_1$ and $\boldsymbol{r}_2$:
\begin{equation}
\label{eq34}
\boldsymbol{\xi}=\frac{L_2\boldsymbol{r}_1-L_1\boldsymbol{r}_2}{L_2-L_1}, \hspace{3mm} \hat{\boldsymbol{v}}=\frac{\boldsymbol{r}_2-\boldsymbol{r}_1}{L_2-L_1}
\end{equation}

Since the sub-network $\mathcal{S}$ is assumed to be rigid, all of its possible representations would differ only in a constant translation $T$ and an arbitrary rotation by a unitary matrix $U$. Assume that another possible representation of $\mathcal{S}$ is denoted by $\tilde{\mathcal{S}}$, which is related to the primary representation through the following equation:
\begin{gather}
\label{eq35}
\tilde{\mathcal{S}}=U\mathcal{S}+T \\
T\in \mathcal{R}^K, \hspace{2mm} U\in \mathcal{R}^{K\times K} \nonumber
\end{gather}

Therefore, new positions for the shared samples can be computed as $\tilde{\boldsymbol{r}}_1=U\boldsymbol{r}_1+T$ and $\tilde{\boldsymbol{r}}_2=U\boldsymbol{r}_2+T$ since both data samples are shared with $\tilde{\mathcal{S}}$.

In the new representation, orientation and position of $\mathcal{L}$ would also confront some transformations. Assume that the new line properties are denoted  by $\tilde{\boldsymbol{\xi}}$ and $\tilde{\hat{\boldsymbol{v}}}$. Based on the relations in (\ref{eq34}) we can compute the new parameters of the line $\tilde{\mathcal{L}}$ through the following equations:
\begin{gather}
\label{eq36}
\tilde{\boldsymbol{\xi}}=\frac{L_2\tilde{\boldsymbol{r}}_1-L_1\tilde{\boldsymbol{r}}_2}{L_2-L_1}=U\left(\frac{L_2\boldsymbol{r}_1-L_1\boldsymbol{r}_2}{L_2-L_1}\right)+T
\\
\tilde{\hat{\boldsymbol{v}}}=\frac{\tilde{\boldsymbol{r}}_2-\tilde{\boldsymbol{r}}_1}{L_2-L_1}=U\left(\frac{\boldsymbol{r}_2-\boldsymbol{r}_1}{L_2-L_1}\right) \nonumber
\end{gather}

Relations in (\ref{eq36}) declare that the starting point of $\tilde{\mathcal{L}}$ has been translated by the constant vector $T$, and both the starting point and the direction vector have been affected by the unitary matrix $U$. This implies that all the samples in the line $\mathcal{L}$ will confront the same transformation as the data samples in $\mathcal{S}$. Therefor the connected line is rigid with respect to the sub-network and the two objects can be merged to form a larger rigid sub-network.

\section{Analytical Solution of Path Mapping Optimization Problem}
In this section an analytical solution for the optimization problem introduced in section 5 of the submitted paper is derived. The optimization problem is formulated as follows:
\begin{align}
(\boldsymbol{\xi}^*,\hat{\boldsymbol{v}}^*)=\argmin_{(\boldsymbol{\xi},\hat{\boldsymbol{v}})}J\left(\boldsymbol{\xi},\hat{\boldsymbol{v}}\right) \nonumber
\end{align}
\begin{equation}
\label{eq_opt}
\subjectto \hspace{2mm} \sum_{k=1}^{K}\left(\hat{v}_p^{(k)}\right)^2=1, \hspace{3mm} p=1,2,...,P
\end{equation}
where:
\begin{align}
\label{eq_J}
J\left(\boldsymbol{\xi},\hat{\boldsymbol{v}}\right) = \frac{1}{2}\sum_{k=1}^{K}\sum_{q=1}^{Q}\left[\frac{1}{m_q}\sum_{i=1}^{m_q}\left(\xi_{\eta_i^{(q)}}^{(k)}+l_{\eta_i^{(q)}}^{(k)}\hat{v}_{\eta_i^{(q)}}^{(k)}\right)^2-\right.
\nonumber \\
\left.\left(\frac{1}{m_q}\sum_{i=1}^{m_q}\left(\xi_{\eta_i^{(q)}}^{(k)}+l_{\eta_i^{(q)}}^{(k)}\hat{v}_{\eta_i^{(q)}}^{(k)}\right)\right)^2\right]
\end{align}

In order to solve the optimization problem in (\ref{eq_opt}), the Lagrangian function corresponding to the objective function $J$ and the set of quadratic constraints in (\ref{eq_opt}) should be formulated:
\begin{align}
\label{eq37}
\mathcal{L}\left(\boldsymbol{\xi}_1,\boldsymbol{\xi}_2,...,\boldsymbol{\xi}_P,\hat{\boldsymbol{v}}_1,\hat{\boldsymbol{v}}_2,...,\hat{\boldsymbol{v}}_P\right)=
\nonumber \\
J\left(\boldsymbol{\xi},\hat{\boldsymbol{v}}\right)-\frac{1}{2}\sum_{p=1}^{P}\lambda_p\left(\sum_{k=1}^{K}\left(\hat{v}_p^{(k)}\right)^2-1\right)
\end{align}

$\lambda_p$s represent the Lagrange multipliers of the optimization. Finding the minimizers of (\ref{eq_J}), $\left(\boldsymbol{\xi}^*,\hat{\boldsymbol{v}}^*\right)$, requires the KKT conditions to be satisfied. First, derivatives of $\mathcal{L}\left(\boldsymbol{\xi},\hat{\boldsymbol{v}}\right)$ with respect to all the variables in (\ref{eq_J}) should become zero. Second, the equality constraints in (\ref{eq_opt}) must hold. Derivatives with respect to $\xi_p^{(k)}$s and $\hat{v}_p^{(k)}$s may be computed as follows:
\begin{align}
\label{eq38}
\frac{\partial\mathcal{L}}{\partial\xi_p^{(k)}}=\sum_{q=1}^{Q}\left(\sum_{\forall i|\atop \eta_i^{(q)}=p}\frac{1}{m_q}\left(\xi_p^{(k)}+l_i^{(q)}\hat{v}_p^{(k)}\right)\right)
\nonumber \\
-\sum_{\forall q|\exists i\atop \Rightarrow \eta_i^{(q)}=p}\frac{1}{m_q^2}\left(\sum_{i=1}^{m_q}\left(\xi_{\eta_i^{(q)}}^{(k)}+l_i^{(q)}\hat{v}_{\eta_i^{(q)}}^{(k)}\right)\right)
\end{align}

\begin{align}
\label{eq39}
\frac{\partial\mathcal{L}}{\partial\hat{v}_p^{(k)}}=\sum_{q=1}^{Q}\left(\sum_{\forall i|\atop \eta_i^{(q)}=p}\frac{l_i^{(q)}}{m_q}\left(\xi_p^{(k)}+l_i^{(q)}\hat{v}_p^{(k)}\right)\right)
\nonumber \\
-\sum_{\forall q|\exists i=i_0\atop \Rightarrow \eta_{i_0}^{(q)}=p}\frac{l_{i_0}^{(q)}}{m_q^2}\left(\sum_{i=1}^{m_q}\left(\xi_{\eta_i^{(q)}}^{(k)}+l_i^{(q)}\hat{v}_{\eta_i^{(q)}	}^{(k)}\right)\right)-\lambda_p\hat{v}_p^{(k)}
\end{align}

Derivations with respect to $\lambda_p$s result in the same constraints in (\ref{eq_opt}). Based on the previous discussions, the preferred low-dimensional representation may be obtained by solving the following set of equations:
\begin{equation}
\label{eq40}
\left\{
  \begin{array}{l l}
    \frac{\partial\mathcal{L}}{\partial\xi_p^{(k)}}=0 \\
    \\
    \frac{\partial\mathcal{L}}{\partial\hat{v}_p^{(k)}}=0
  \end{array} \right.\
  \begin{array}{l l}
  	p=1,2,...,P \\
  	k=1,2,...,K  
  \end{array}
\end{equation}

And also satisfying following equality constraints:
\begin{equation}
\label{eq41}
\sum_{k=1}^{K}\left(\hat{v}_p^{(k)}\right)^2=1, \quad p=1,2,...,P
\end{equation}

In order to simplify the notations, the linear equations in (\ref{eq40}) can be rewritten in the following matrix form:
\begin{equation}
\label{eq42}
\left\{
  \begin{array}{l l}
    A\boldsymbol{\xi}+B\hat{\boldsymbol{v}}=0 \\    
    A'\boldsymbol{\xi}+B'\hat{\boldsymbol{v}}=\Lambda\hat{\boldsymbol{v}}
  \end{array} \right.\
\end{equation}
where $A, B, A'$ and $B'$ are $P\times P$ matrices whose entries can be calculated by the formulations given in (\ref{eq14}),(\ref{eq15}),(\ref{eq16}) and (\ref{eq17}) respectively. $\Lambda$ is a $P\times P$ diagonal matrix consisting of $\lambda_p$s on its main diagonal ($\Lambda_{p,p}=\lambda_p, p=1,2,...,P$).

	
Equations in (\ref{eq42}) will lead to a non-linear matrix equation as follows:
\begin{gather}
\label{eq43}
\hspace{3mm} \left(B'-A'A^{\dagger}B\right)\hat{\boldsymbol{v}}^*=\Lambda\hat{\boldsymbol{v}}^* \\
\boldsymbol{\xi}^*=-A^{\dagger}B\hat{\boldsymbol{v}}^* \nonumber
\end{gather}

$A^\dagger$ is the pseudo-inverse of the matrix $A$. The matrix $A$ is singular and thus non-invertible since each one of its rows sum to zero. This property can be investigated from (\ref{eq_A_mat}). Therefore $A$ has an all-one $\left(\boldsymbol{1}_{P\times 1}\right)$ eigenvector with a zero eigenvalue:
\begin{equation}
\label{all_one_eigenvector}
A\left(\boldsymbol{1}_{P\times 1}\right)=0
\end{equation} 

The all-one eigenvector in $A$ models a constant translation in the position of all the data samples in low-dimensional representation (Note that $A$ is associated with the optimal starting points $\boldsymbol{\xi}^*$ in (\ref{eq42})). As will be discussed in section 2 of this document this translation does not affect the proximity information. Placement of $A^{\dagger}$ instead of the true inverse $A^{-1}$ in (\ref{eq43}) forces the whole low-dimensional representation to be centered on the origin. The same story holds for $B$ (since it is singular too), so that its null eigenvector yields the ``rotation" invariance property of parameters in mathematical terms.

Before going any further with (\ref{eq43}), let us investigate if there could be any alternative way to reach a straight forward solution for the original problem in section V of the submitted paper. The main reason for the constraints in (\ref{eq41}) is to avoid any trivial solution in which all the variables become zero. This goal could be gained by imposing different constraints in the optimization problem in (\ref{eq_opt}). The new set of constraint inequalities can be formulated as follows:
\begin{equation}
\label{eq44}
\sum_{p=1}^{P}\left(\hat{v}_p^{(k)}\right)^2\ge1, \hspace{3mm} k=1,2,...,K
\end{equation}

The above set of inequalities although represents a whole different mathematical meaning, however, results into a small difference in the formulation given in (\ref{eq43}):
\begin{gather}
\label{eq45}
\hspace{3mm} \left(B'-A'A^{\dagger}B\right)\hat{\boldsymbol{v}}^*=\hat{\boldsymbol{v}}^*\Lambda \\
\boldsymbol{\xi}^*=-A^{\dagger}B\hat{\boldsymbol{v}}^* \nonumber
\end{gather}

Also, inequality constraints in the new formulations impose extra KKT conditions, which are:
\begin{equation}
\label{eq_new_con}
\left\{
\begin{array}{l l}
\lambda_p\ge0 \\
\lambda_p\left[\sum_{q=1}^{P}\left(\hat{v}_q^{(k)}\right)^2-1\right]=0
\end{array} \right.\
\quad p=1,2,...,P
\end{equation}
where $\lambda_p$s are called KKT multipliers in this formalism.

Unlike (\ref{eq43}), matrix equation of (\ref{eq45}) represents an eigenvector problem that can be efficiently solved by a polynomial-time algorithm. Therefore the problem of finding appropriate low-dimensional representations for a high-dimensional dataset results into eigen-decomposition of a $P\times P$ matrix.

We have denoted the square $P\times P$ matrix $B'-A'A^{\dagger}B$ in (\ref{eq45}) with $\psi$ in section V. Again it can be shown that $\psi$ is a symmetric semi-definite matrix with real and non-negative eigenvalues. This will satisfy the constraints in (\ref{eq_new_con}). Also the eigenvector computation of $\psi$ can be accomplished by a singular value decomposition (SVD) problem with up to $O(P^3)$ operations.
 
Substitution of (\ref{eq45}) into (\ref{eq_J}) reveals that the appropriate columns of $\hat{\boldsymbol{v}}^*$ matrix which minimize the objective function are $K$ eigenvectors of $\psi$ that correspond to the $K$ smallest positive eigenvalues. Note that the $K$ smallest eigenvalues of $\psi$ are the minimum KKT multipliers of the optimization problem. 

Finally we should apply an appropriate normalization step to match the results with the set of constraint in (\ref{eq_opt}). The final solution of the optimization problem can be formulated as follows:
\begin{align}
\label{eq46}
\psi{U}=U{\Lambda}, \hspace{3mm} U=\left[\boldsymbol{u}_1|\boldsymbol{u}_2|...|\boldsymbol{u}_P\right]
\nonumber \\
\boldsymbol{u}_p\in \mathcal{R}^{P\times 1}, \hspace{3mm} p=1,2,...,P
\nonumber \\
\nonumber \\
\hat{\boldsymbol{v}}^{(k)}_p=\frac{u^{(p)}_{P-\left(k+k_0\right)+1}}{\sqrt{\sum_{k=1}^{K}\left(u^{(p)}_{P-\left(k+k_0\right)+1}\right)^2}}
\end{align}
where $\boldsymbol{u}_i$s are columns of the unitary matrix $U$ obtained from a singular value decomposition step discussed before. $u_i^{(p)}$ is the $p$th component of the $\boldsymbol{u}_i$ vector. It has been assumed that the eigenvalues and corresponding eigenvectors of the $U$ matrix are in descending order, so the last eigenvector corresponds to the smallest eigenvalue. $k_0$ is the number of zero eigenvalues. Based on the normalization performed in (\ref{eq46}) the direction vectors (rows of $\hat{\boldsymbol{v}}^*$) will be normal vectors with unit lengths. Optimal starting points can be obtained using (\ref{eq45}).

\ifCLASSOPTIONcompsoc
  \section*{Acknowledgments}
\else
  \section*{Acknowledgment}
\fi

The authors would like to thank M. F. Azampour and A. Ghafari for their helpful comments throughout the paper.

\ifCLASSOPTIONcaptionsoff
  \newpage
\fi



%
{\small
\input{bare_jrnl_compsoc.bbl}

}

\begin{IEEEbiography}{Amir Najafi}
\end{IEEEbiography}

\begin{IEEEbiography}{Amir Joudaki}
\end{IEEEbiography}


\begin{IEEEbiography}{Emad Fatemizadeh}
\end{IEEEbiography}




\end{document}